# The manipulation of massive ro-vibronic superpositions using time-frequency-resolved coherent anti-Stokes Raman scattering (TFRCARS): from quantum control to quantum computing

*R. Zadoyan, D. Kohen,[†] D. A. Lidar,[§] V. A. Apkarian, Department of Chemistry, University of California, Irvine CA 92697-2025.*

## ABSTRACT

Molecular ro-vibronic coherences, joint energy-time distributions of quantum amplitudes, are selectively prepared, manipulated, and imaged in Time-Frequency-Resolved Coherent Anti-Stokes Raman Scattering (TFRCARS) measurements using femtosecond laser pulses. The studies are implemented in iodine vapor, with its thermally occupied statistical ro-vibrational density serving as initial state. The evolution of the massive ro-vibronic superpositions, consisting of $10^3$ eigenstates, is followed through two-dimensional images. The first- and second-order coherences are captured using time-integrated frequency-resolved CARS, while the third-order coherence is captured using time-gated frequency-resolved CARS. The Fourier filtering provided by time integrated detection projects out single ro-vibronic transitions, while time-gated detection allows the projection of arbitrary ro-vibronic superpositions from the coherent third-order polarization. A detailed analysis of the data is provided to highlight the salient features of this four-wave mixing process. The richly patterned images of the ro-vibrational coherences can be understood in terms of phase evolution in rotation-vibration-electronic Hilbert space, using time circuit diagrams. Beside the control and imaging of chemistry, the controlled manipulation of massive quantum coherences suggests the possibility of quantum computing. We argue that the universal logic gates necessary for arbitrary quantum computing – all single qubit operations and the two-qubit controlled-NOT (CNOT) gate – are available in time resolved four-wave mixing in a molecule. The molecular rotational manifold is naturally "wired" for carrying out all single qubit operations efficiently, and in parallel. We identify vibronic coherences as one example of a naturally available two-qubit CNOT gate, wherein the vibrational qubit controls the switching of the targeted electronic qubit.

---

[†] Present address: Smith College, Chemistry Department, Northampton, MA 01063.
[§] Permanent address: Chemistry Department, University of Toronto, 80 St. George St., Toronto, Ontario, Canada M5S 3H6.

# I. INTRODUCTION

Coherent anti-Stokes Raman scattering (CARS) is a well established and broadly applied spectroscopic tool,[1,2] the subject of textbooks.[3,4,5] When carried out with ultrafast lasers, within a single experiment, CARS combines the elements of preparation, manipulation, and interrogation of molecular coherences. These are the key ingredients of quantum control, be it applied to quantum chemistry,[6,7,8] or quantum computing.[9,10] It is from this perspective that we present our multi-dimensional time- and frequency-resolved CARS studies on the well-characterized system of diatomic iodine in the gas phase, and at room temperature.

Quantum coherences can be defined by their two-dimensional mapping along relevant conjugate variables, such as time and energy.[1] This is familiar in practice in the characterization of optical pulses by such means as frequency resolved optical gating (FROG).[11] An equivalent characterization of molecular coherences is enabled through joint time-frequency-resolved CARS (TFRCARS) spectroscopy. This, we recently demonstrated through time-gated, frequency-resolved detection of the anti-Stokes polarization (TGFCARS) in iodine vapor at room temperature.[12] The experiment served to illustrate that the third-order coherence of the molecule could be imaged directly as an interferogram on the time-frequency plane, and that ro-vibronic coherences consisting of superpositions of $10^3$ states could be rigorously decomposed in such maps. There, we emphasized that although fs pulses are used, the third-order coherence in resonant CARS can be detected with rotational resolution, since it evolves freely until the unset of decoherence. In room temperature vapor iodine ro-vibrational decoherence would be determined by pure dephasing due to the Doppler inhomogeneous width of lines (t ~ $10^{-9}$s). With regard to chemical dynamics, this implies that transients captured in fs time can be analyzed with high resolution ($\Delta\omega$ ~ 0.02 cm$^{-1}$). Here, we extend the analysis to first- and second- order coherences, detected through time-integrated frequency-resolved CARS. As in our prior development, we maintain the intuitive languages of discussing vibrational coherences in terms of wavepackets,[13] while ro-vibrational coherences[14,15] are treated as phase evolution in vector space. As an example, in such a treatment, the counter-intuitive result that condensed phase CARS signals should be more deeply modulated than in the gas phase can be understood in terms of the spatial extent of vibrational superpositions.[16] In contrast, the description of time evolution of rotational superpositions as purely phase evolution of eigenstates in Hilbert space is more natural and suggestive in computational applications.

---

[1] In keeping with common practice energy, frequency, wavelength and angular frequency are interchangeably used in the text. The time-energy space is invariably refereed to as time-frequency while the experimental data are presented on a time-wavelength axis.

The concept that the controlled manipulation of quantum coherences can be used for quantum computation or information transfer, dates to an annunciation on general grounds by Feynman,[17] followed by the formal arguments presented by Deutsch.[18] This very active field of research was more recently catapulted forward by the development of algorithms that take advantage of quantum parallelism to achieve exponential speed-up in particular tasks. The most notable among these being Shor's algorithm for factorization of a number into its primes,[19] and Grover's search algorithm,[2] to search an arbitrarily large data base with a single query.[20] The field has since advanced dramatically, with physical proofs of principle based on the demonstration of universal quantum logic gates, realized through a variety of experimental approaches. Among examples are: all optical interferometry,[21,22] cavity quantum electrodynamics,[23] ion traps,[24] superconducting Josephson junctions,[25] and an explosion of studies in NMR following the initial propositions.[26,27] More specific tasks, such as a search algorithm through the preparation and manipulation of Rydberg Superposition states in atomic beams,[28] have also been demonstrated. The latter belongs to the category that does not require entanglement,[29] and therefore can be argued that does not allow exponential speed-up without the expenditure of an exponential overhead in resources.[30] Physical realizations that involve entanglement,[31] to-date, have been limited to several qubits (the quantum analog of a bit),[32] with extensions to larger numbers being a nontrivial challenge. The multidimensional coherent spectroscopy of molecular ro-vibronic coherences involves the controlled manipulation of massive Quantum superpositions; as such, it presents an opportunity for executing quantum logic on a very large manifold of states. After presenting the experimental data and their interpretation, we discuss the natural structure of the tensor space in which the ro-vibronic polarization is manipulated, and the naturally available "wiring" of quantum logic gates for computational tasks. The massiveness of the ro-vibrational superposition that is manipulated, consisting of $10^3$ states, and the precision with which coherences can be transferred between ro-vibronic states, makes the system particularly attractive for computation with massive parallelism.

Molecular iodine is chosen for these studies because of its spectroscopic convenience, well characterized spectroscopy, and the fact that it has already been scrutinized by fs CARS experiment and theory in the gas phase.[33,34,35,36] Although unintended, we discover several unexpected electronic scattering channels in iodine. We discuss these for completeness. The multiple electronic resonances that lead to complexity in the data can add significant flexibility in the tailored manipulation of electronic coherences.

Prior to presenting the experimental results and their analysis, we establish the frameworks to be used in the interpretation of vibronic and ro-vibronic contributions to resonant CARS, along lines already introduced in two preceding papers.[12,16] Coherent Raman scattering (CRS) is a four-wave mixing process that involves measuring the third-order material polarization in response to the application of three input beams.[5] The fourth rank hyperpolarizability tensor that mediates the process allows selectivity by experimental control on time-orderings and Cartesian components of the applied fields, hence the proliferation of acronyms devised to highlight specialization. In fs CARS, input pulses of different color are chosen such that the detected coherent material polarization occurs to the blue of the input pulses. Single color fs CRS, often identified by its frequency domain acronym of degenerate four-wave mixing (DFWM), has been implemented in small gas phase molecules to explore aspects of molecular control.[37,38] In condensed media, the same measurements appear under the heading of three-pulse photon echo.[39,40] Both time-gated[41] and frequency-resolved[42,43] stimulated photon echo measurements have been performed, which combined,[44] would be equivalent to our TGFCARS experiments. The more complete four-wave mixing measurements involve heterodyne detection of the radiation, to yield amplitudes and phases of all fields.[45,46] Heterodyne detected three-pulse photon echo measurements have been extended to the infrared, and have been implemented to the study of peptides.[47] Reviews of these inherently multi-dimensional fs spectroscopies of vibronic excitations have recently appeared.[48,49]

---

[2] Although provably faster, Grover's algorithm gives square root speed-up over classical algorithms.

### A) Vibronic Coherences

In time resolved CARS, the third-order polarization, $P^{(3)}$, induced by the application of three short laser pulses is probed.[3-5] As in the more common implementations, here too, two of the pulses are chosen to have the same color and are identified as pump, P and P'; while the third, red-shifted pulse is identified as Stokes, S. The signal consists of the polarization propagating along the anti-Stokes wavevector, $\mathbf{k}_{AS} = \mathbf{k}_p + \mathbf{k}_{p'} - \mathbf{k}_S$. Choosing both pump and Stokes colors to overlap the dipole allowed $X(^1\Sigma_{0_g^+}) \leftrightarrow B(^3\Pi_{0_u^+})$ electronic transition, we may expect the rotating wave approximation to hold, and the consideration may be limited to scattering processes that are resonant in all orders. We will initially restrict the consideration to the two-electronic state molecular Hamiltonian,

$$H = |X\rangle H_X \langle X| + |B\rangle (H_B + T_e)\langle B| \qquad (1)$$

in which $H_B$ and $H_X$ are the vibrational Hamiltonians in the excited and ground electronic states, respectively. Within the interaction representation of time dependent perturbation theory, $P^{(3)}$ is the expectation value of the dipole operator:[50]

$$P^{(3)}(t) = \langle \varphi_X^{(0)}(t) | \hat{\mu} | \varphi_B^{(3)}(t) \rangle + \langle \varphi_X^{(2)}(t) | \hat{\mu} | \varphi_B^{(1)}(t) \rangle + c.c. \tag{2}$$

where

$$\hat{\mu} = \mu(|\varphi_X\rangle\langle\varphi_B| + |\varphi_B\rangle\langle\varphi_X|) \tag{3}$$

and

$$|\varphi^{(n)}(t)\rangle = \frac{i}{\hbar} \int_{-\infty}^{t} dt'\, e^{-iH(t-t')} \hat{\mu} \cdot \mathbf{E}(t') |\varphi^{(n-1)}(t')\rangle \tag{4}$$

with applied fields described by their envelopes $E_l(t)$:

$$\mathbf{E}_l(t) = \hat{\varepsilon}_l [E_l(t) e^{-i\omega_l t} + E_l^*(t) e^{+i\omega_l t}] \qquad \text{where } l = P, P', S \tag{5}$$

The possible contributions to (2), which are generated by the permutations of the P, P', and S pulses in (5), can be illustrated by the familiar double-sided Feynman diagrams, as in Figures 1 and 2. The equivalent time-circuit diagrams, which prove useful in describing the material response in both state representation and in classical[51] or semiclassical propagations,[52] are also shown in the figures. As it will become clearer, the time-circuit diagrams are quite useful in keeping track of the material response in state space.

Choosing $\omega_P$ to the red of the absorption maximum and $\omega_S$ outside the absorption band ($\hbar\omega_p - \hbar\omega_s > k_B T$), only the first term in Eq. 2 (Fig. 1) can be electronically resonant in all three pulses. Further, by experimentally choosing the sequence P, followed by S, followed by P', we may transcribe the diagram in Fig. 1 to the explicit third-order perturbation expression:[13]

$$\begin{aligned}
P^{(3)}_{\mathbf{k}_{AS}}(t) &= P^{(0,3)}_{\mathbf{k}_{AS}}(t) + P^{(3,0)}_{\mathbf{k}_{AS}}(t) \\
&= \frac{k}{\hbar^3} \int_{-\infty}^{t} dt_3 \int_{-\infty}^{t_3} dt_2 \int_{-\infty}^{t_2} dt_1\, e^{-i(\omega_{P'} - \omega_S + \omega_P)t} \\
&\times \langle \varphi_X e^{iH_X t/\hbar} | \hat{\mu} | e^{-iH_B(t-t_3)/\hbar} \hat{\mu} E_{P'}(t_3) e^{-iH_X(t_3-t_2)/\hbar} \hat{\mu} E_S^*(t_2) e^{-iH_B(t_2-t_1)/\hbar} \hat{\mu} E_P(t_1) e^{-iH_X t_1/\hbar} \varphi_X \rangle \\
&+ c.c.
\end{aligned} \tag{6}$$

The content of (6) can be readily visualized in the wavepacket picture of Fig. 3. For all times, the bra state vector $\langle \varphi_X^{(0)}(t) |$ evolves subject to the bare, ground state molecular Hamiltonian. The fields act on the ket state. At $t_1$, the pump prepares a wavepacket in the B state, $|\varphi_B^{(1)}(t)\rangle$, in the Franck-Condon window carved out by the pump laser. In this first order coherence, the packet on the B potential evolves until $t = t_2$, when the Stokes pulse arrives. The portion of the packet that overlaps with the Stokes window can now be transferred to prepare the second-order (or the Raman) packet, $|\varphi_X^{(2)}(t)\rangle$. The Raman packet evolves on the X state until $t = t_3$, when the P' pulse acts. Now amplitude proportional to overlap of the Raman packet with the pump window is transferred to the B-state, to prepare the third-order packet $|\varphi_B^{(3)}(t)\rangle$. In this third-order coherence, the system evolves freely, radiating every time the vibrational packet reaches the anti-Stokes window, which is located at the inner turning point of the B-state potential where the energy conservation condition $\delta[\omega_{AS} - (2\omega_P - \omega_S)]$ can be satisfied. This resonantly created third-order polarization will persist long after the termination of pulses, until destroyed by collisions. Recursions in the third-order polarization imply structure in the AS spectrum, as already argued and demonstrated.[12] If we make an analogy with photon echo, the observable recursions in the third-order polarization would be more appropriately described as *photon reverberations.* Note, in first- and third-order the bra and ket are in separate electronic states, therefore the system is in a ro-vibronic (rotation-vibration-electronic) coherence. In second-order, the Raman packet is in a ro-vibrational coherence in an electronic population on the X-state. Finally, note that the complex conjugation indicated in (6) ensures that the third-order polarization is real, and that a diagram conjugate to each of those illustrated in Figures 1 and 2 is operative in the process. The concepts of wavepackets and coherences derive from somewhat different starting points; nevertheless, we marry the languages to take advantage of the intuition contained in each.

### B) Ro-vibrational Coherences:

The complete rotation-vibration contribution to $P^{(3)}$ can be given in terms of the density matrix in Hilbert space:

$$\rho(t) = \sum_{\chi',v',j'} \sum_{\chi,v,j} \sum_{m'=-j'}^{j'} \sum_{m=-j}^{j} c(\chi',\chi,v',v,j',j,m',m) |\chi',v',j',m';t\rangle\langle\chi,v,j,m;t| \qquad (7a)$$

in which $\chi$, $v$, $j$, $m$ designate electronic, vibrational, rotational and magnetic quantum numbers of eigenstates ($\chi,\chi'$ = X, B). After three interactions with the laser field, the expectation value of the dipole over the third-order polarization $Tr[\hat{\mu}\rho^{(3)}(t)]$ is measured. Starting with a given initial eigenstate of the thermal density $|X,v,j\rangle\langle X,v,j|$, the dipole operator (3) ensures that after the first interaction with the radiation field the first-order coherence is prepared:

$$\rho^{(1)}(t) = \sum_{v',v,j} \left[ c(v,v',j+1,j)|B,v',j+1;t\rangle\langle X,v,j| + c(v,v',j-1,j)|B,v',j-1;t\rangle\langle X,v,j| \right] + c.c.$$

$$= |\varphi_B^{(1)}(t)\rangle\langle\varphi_X^{(0)}(t)| + |\varphi_B^{(1)}(t)\rangle\langle\varphi_X^{(0)}(t)| + c.c.$$

(7b)

in which the state coefficients are a function of the applied field (4), and contain Frank-Condon factors and rotational matrix elements. The second pulse prepares the electronic population, $\rho^{(2)}(t) = |\varphi_X^{(2)}(t)\rangle\langle\varphi_X^{(0)}(t)| + c.c.$, which is a ro-vibrational coherence $\rho^{(2)}(t) = \sum c(v'',j'')|X,v'',j'';t\rangle\langle X,v,j;t| + c.c.$ with $j'' = j, j\pm 2$ and $v'' \neq v$ since P and S pulses do not have any spectral overlap. Finally, according to Fig. 1, in the four-wave process a given eigenstate is forward propagated on excited B→X→B states over the intervals $t_{21}\to t_{32}\to t_{43}$, then dipole projected back onto the original eigenstate and propagated in reverse-time over the $t_{34}$ interval. This three-time correlation contributes a unity (reaches its maximum value) when the phase accumulated over the time-circuit is an integer multiple of $2\pi$:[3]

$$\begin{aligned}\Omega(t;',",''') &= [E_{v,j}^X(t_1 - t_4) + E_{v''',j'''}^B(t_4 - t_3) + E_{v'',j''}^X(t_3 - t_2) + E_{v',j'}^B(t_2 - t_1)]/\hbar \\ &= \omega_{v,j}^X t_{41} + \omega_{v''',j'''}^B t_{34} + \omega_{v'',j''}^X t_{23} + \omega_{v',j'}^B t_{12} \\ &= 2\pi n\end{aligned} \quad (8)$$

This defines the condition for phase coherence in CARS, for a given path in state space. Note, for a given path in vibrational space – *v, v', v", v'''* – the *Δj=±1* dipole selection rules lead to six rotational paths to close the time-circuit. This is illustrated in Fig. 4, in a standard diagram showing optical transitions with vertical arrows (Fig. 4a), and in a schematic "wiring" diagram (Fig. 4b). The bra-state, *<v,j;t|*, represented by the lower line in Fig. 4b, acts as a reference for defining instantaneous phases of the evolving ket-states. Thus, the instantaneous complex amplitudes in the coherent superposition *a'|v',j+1><v,j| + b'|v',j-1><v,j|* prepared by the P-pulse are given as: $a'(t) = c'e^{-i\Omega'(t)}$ with $\Omega(t) = (\omega_{v',j+1}^B - \omega_{v,j}^X)(t - t_1)$, where *c'* is determined by the laser field and transition matrix elements. For a laser pulse short in comparison to ro-vibrational periods, phase evolution under the pulses may be ignored. Thus, the amplitudes in the final superposition *a'''|v''',j+1><v,j| + b'''|v''',j-1><v,j|* are controlled by the interference between the three paths that connect each of the final *j±1* pair of eigenstates to the initial coherent *j±1* pair, with path lengths controlled by the delay between laser pulses. The detectable polarization consists of contributions from closed circuits from each of the thermally occupied initial eigenstates, as a squared, real quantity with a definite phase. Since it is the bulk polarization that is detected experimentally, and the individual time-circuits starting from each statistical initial state is phase-locked by the sudden action of the P-pulse at *t = t₁*, all components of the radiation may interfere among themselves. Both phase and magnitude information is contained in the polarization, and can be retrieved[46] (phase, to within a sign in the present measurements). We will discuss the information stored, manipulated, and retrieved in the various order quantum coherences after presenting the experiment and its analysis.

---

[3] Contrary to the practice of identifying the ground state quantum numbers with double primes, we will identify the ground state indices without primes while first, second and third order states are identified with one, two, and three primes, respectively.

## II. EXPERIMENTAL

Experimentally, it is more natural to think about the four-wave mixing process in terms of bulk polarization.[3] In the long-wave limit, the third-order bulk susceptibility probed in CARS, $P^{(3)}$, results from molecular contributions, $P^{(3)} = NP^{(3)}$, where $N$ is the molecular number density, and the third-order molecular polarization is the response to three applied fields mediated by the fourth rank hyperpolarizability tensor, $\gamma$:

$$\mathbf{P}_\rho^{(3)}(\omega_0,t) = \sum_{\sigma\tau\nu}\sum_{(1,2,3)} \gamma_{\rho\sigma\tau\nu}(-\omega_0,\omega_1,\omega_2,\omega_3)\mathbf{E}_\sigma(\mathbf{k}_3\omega_3,t_3)\mathbf{E}_\tau(\mathbf{k}_2\omega_2,t_2)\mathbf{E}_\nu(\mathbf{k}_1\omega_1,t_1) \qquad (9)$$

We use two colors for the three input beams: $\omega_1 = \omega_3 \equiv \omega_P = \omega_{P'}$ and $\omega_2 \equiv \omega_S$, in the forward BOXCARS arrangement illustrated in Figure 5.[53] The three beams propagating along separate wavevectors, after passing through separate delay lines, are brought into focus using a single achromatic doublet (fl = 25 cm). Two pinholes serve to spatially filter, $P_{\mathbf{k}_{AS}}^{(3)}(t) = \int d\mathbf{r} e^{i(\mathbf{k}_P + \mathbf{k}_{P'} - \mathbf{k}_S)\mathbf{r}}\mathbf{P}^{(3)}$, the forward-scattered coherent polarization along the anti-Stokes (AS) wavevector:

$$\mathbf{k}_{AS} = \mathbf{k}_p + \mathbf{k}_{p'} - \mathbf{k}_S \qquad (10a)$$

with associated energy conservation condition:

$$\omega_0 \equiv \omega_{AS} = 2\omega_P - \omega_S \qquad (10b)$$

In conventional implementations, the signal consists of the directional coherent AS polarization collected with a single element square-law detector, as a function of delay between pulses:

$$S(t_1,t_2,t_3) = \int dt_4 [P_{\mathbf{k}_{AS}}^{(3)}(t_4;t_3,t_2,t_1)]^2 \qquad (11)$$

Instead, we record the spectrally resolved AS radiation in one of two modes: time-integrated or time-gated detection.

*In Time-integrated frequency-resolved CARS (TIFRCARS)*, the spatially filtered AS polarization is dispersed in a 1/4-m monochromator and detected using a CCD array. The integration is over $t \equiv t_4$, and since pulsed lasers are used, the limits of integration can be extended to $\pm\infty$:

$$I_{AS}(\omega;t_1,t_2,t_3) = [\int_{-\infty}^{\infty} dt e^{-i\omega t} P_{AS}^{(3)}(t)]^2 \qquad (12)$$

The recorded AS spectrum is composed of the signal elements (pixels) of the array

$$S(\omega_0) = [\int d\omega f(\omega-\omega_0,\delta\omega) \int_{-\infty}^{\infty} dt e^{-i\omega t} P_{AS}^{(3)}(t)]^2$$
$$= [\int d\omega f(\omega-\omega_0,\delta\omega) P_{AS}^{(3)}(\omega)]^2 \qquad (13)$$

with $f(\omega-\omega_0,\delta\omega)$ defined as the amplitude bandpass of the spectrometer. By selecting the time delays between pulses, $t_1$, $t_2$, $t_3$, the time integrated spectrum (12) yields images of the first and second coherences. Thus, following the time-circuit diagram in Fig. 1, the AS spectrum obtained as a function of $t_{21}$ with $t_{32} = 0$ (i.e., with P pulse preceding the coincident S and P' pulses), yields the evolution in *first-order coherence*:

$$S(\omega,t_{21}) = S(\omega;\delta[t_3-t_2],t_2-t_1) \equiv S(\omega,t_<)$$
$$= [\int_{-\infty}^{\infty} dt e^{-i\omega t} P_{AS}^{(3)}(t,\delta[t_3-t_2],t_2-t_1)]^2 \qquad (14a)$$

Similarly, the time integrated spectrum obtained with coincident P and S pulses, $\delta[t_2-t_1]$, as a function of delay of the P' pulse, yields the evolution in *second–order coherence*:

$$S(\omega,t_{32}) = S(\omega;t_3-t_2,\delta[t_2-t_1]) \equiv S(\omega,t_>)$$
$$= [\int_{-\infty}^{\infty} dt e^{-i\omega t} P_{AS}^{(3)}(t,t_3-t_2,\delta[t_2-t_1])]^2 \qquad (14b)$$

Since P and P' pulses are identical, in practice, one of the pump pulses is overlapped in time with the Stokes pulse and the second pump pulse is scanned in time. The experimental convention uses negative time, $S(t_<)$, to identify measurement in first-order coherence over $t_{21}$, and positive time signal, $S(t_>)$, to identify measurement of evolution in second-order coherence, over $t_{32}$. Experimentally, coincident pulses imply two pulses with a relative delay that is small in comparison to their width. The relative delay is adjusted while optimizing a particular signal.

*Time-gated frequency-resolved CARS (TGFCARS)* is implemented by passing the AS beam through a Kerr gate before entering the monochromator. The Kerr cell consists of a 5-mm long cuvette filled with $CS_2$, held between a cross-polarized pair of prisms (see Fig. 5). A few percent of the 800 nm fundamental of the Ti:Sapphire laser, polarized at 45° relative to the CARS beam and passed through a separate delay line, is used to induce Kerr rotation in the cell. Spatial and temporal overlap of the gate pulse with the CARS pulse is obtained using the non-resonant signal from air. This also provides a calibration of the gate width, $\delta t$ = 500 fs, in agreement with the known Kerr response of $CS_2$.[54] For a fixed time sequence of input pulses, the CARS spectrum as a function of gate delay, $t_4$, yields the two-dimensional image of the evolving third-order coherence

$$S(\omega_0, t_4) = \int dt\, g(t_4 - t, \delta t) S(\omega_0, t) \qquad (15)$$

Since, the width of the Kerr gate is comparable to vibrational periods in iodine; the signal will smooth over the vibrational modulation. The spectral resolution, $\delta$, is determined by the broader of the spectrometer bandpass or the inverse gate width: $\delta^2 = \delta\omega^2 + c^2 \delta t^2$. TGFCARS is similar to heterodyne detection, however, since the pulses used in the experiment are not phase locked, only an envelope function appears in (15), and phase information is lost.

    The fs laser source used in these experiments consists of a Ti:Sapphire oscillator, which is chirped pulse amplified at 1 kHz to an energy of 700 μJ/pulse, and compressed to a pulse width of 70 fs. The pulse is split with a 50% beam splitter to pump two three-stage optical parametric amplifiers (OPA). The OPA output is frequency up-converted by sum generation, to provide tunability in the 480-2000 nm spectral range. The time-frequency profiles of the pulses are adjusted with two-prism compressors following each OPA. The pulses are not transform-limited. The OPA output is adjusted to yield a bandwidth of 450 cm$^{-1}$.

    Critical alignment of beams in space and time is crucial for the successful execution of the experiments. This is achieved by first optimizing the non-resonant CARS signal obtained from a 200 μm thick glass plate, then removing the plate and optimizing the non-resonant CARS signal from air, and finally inserting the static quartz cell containing $I_2$ heated to 50°C in the beam overlap region. The cell is positioned to optimize the signal, then beam paths are adjusted to compensate for changes introduced by the cell windows.

### III. RESULTS AND ANALYSIS

#### A) The ro-vibrational third-order coherence – TGFCARS:

A time-gated, frequency-resolved CARS (TGFCARS) spectrum is shown in Figure 6. The image was obtained with coincident S+P' pulses, $t_{23}$ =0, delayed from the P pulse by one vibrational period in the B state, $t_{12}$ = 380 fs. A nearly identical image is obtained with all three pulses in coincidence, $t_{12}=t_{23}=0$. The AS polarization, which is patterned by interference among ~$10^3$ ro-vibrational transitions is followed for 55 ps, with an effective resolution of 20 cm$^{-1}$. This direct image of the third-order coherence was previously analyzed in some detail.[12] The main features of the interferogram are easily understood:
  a) An intense signal occurs at *t = 0*, which within 1.5 ps decays to a 12 ps period of silence indicating complete destructive interference. At *t = 0*, since $\Omega = 0$ for all eigenstates in the superposition, all states radiate in phase. Given the dense manifold of ro-vibrational states under the sampling gate, the initial decay is simply the dephasing time given by the inverse of the detection bandpass (20 cm$^{-1}$). The period of silence corresponds to a flat distribution of phases of rotational transitions collected under the time-frequency window of detection.
  b) As the highest observable rotational states execute an $\Omega = 2\pi$ phase rotation, Eq. 8, the polarization reappears, and develops into five chirped trails of rotational revivals in the five vibrational states v =31-36 of the third-order wavepacket. The periodicity of recursions can be understood as the beat between adjacent spectral components of the polarization. The hyperbolic trails result from the inverse dependence of recursion time, $\tau_r$, on rotational quantum number:
$$1/\tau_r = 2(B"-B')j/c \qquad (16)$$
  where B" and B' are the rotational constants in the ground and excited states, respectively.[12]
  c) The trails develop a complex pattern due to interference between different rotational transitions of a single vibrational transition, as the difference between rotational recursions (winding numbers) of rotational states in a given vibration increases with time. Interference also occurs as rotational recurrences from different vibrations overlap in the time-frequency plane.

All details of the two-dimensional interferogram can be reproduced using the accurately known spectroscopic constants of iodine,[55] as shown in the simulation in Fig. 6. The image of the third-order polarization is reconstructed by considering the AS radiation sampled under the time-frequency gate, consisting of P- and R-branch transitions ( *j-1→j* and *j+1→j* ):

$$S(\omega,t_4) = \left\{ \int_{\omega-\delta\omega}^{\omega-\delta\omega} d\omega \int_{t+\delta t}^{t-\delta t} dt\, g(\omega, t-t_4) \sum_{v,v'''} \sum_j e^{-\beta E^X_{v,j}} [c(v,j,v''',j\pm 1) e^{-i(E^B_{v''',j\pm 1} - E^X_{v,j})t_4/\hbar}] \right\}^2 \quad (17)$$

with coefficients:

$$c(v,j,v''',j\pm 1) = \mu_{BX}^4 c(j,j\pm 1)c(v,v''')\delta[E_L - E_{v,j}]^3$$

$$= \mu_{BX}^4 [D_{j,j\pm 1}^2 D_{j\pm 1,j\pm 2}^2 + D_{j,j\pm 1}^4 + D_{j,j\pm 1}^2 D_{j,j\mp 1}^2] \sum_{v',v''} \langle v|v'''\rangle\langle v'''|v''\rangle\langle v''|v'\rangle\langle v'|v\rangle$$

$$\times E(\omega_p)E(\omega_S)E(\omega_p)\delta[\hbar\omega_p - E_{v''',j''',v'',j''}]\delta[\hbar\omega_S - E_{v''j'',v'j'}]\delta[\hbar\omega_p - E_{v'j',vj}]$$

(18a)

where, for linearly polarized all-parallel fields[56]

$$D_{j,j\pm 1} = \sum_{m=-j}^{j} \langle j,m|10|j\pm 1,m\rangle \quad (18b)$$

While their evaluation is straightforward, since within the sampling bandpass the transition coefficients are slow functions of wavelength, we set $c(v,j,v''',j\pm 1) = 1$ in the simulations. The good comparison between simulation and experiment in Fig. 6 validates this approximation, and suggests robustness of the detectable polarization. The latter consideration is quite valuable in the quantum computational applications to be considered.

The interference structure in the third-order polarization occurs on several time-frequency scales, therefore the detected image can vary substantively with the choice of laser colors. The simulations in Fig. 7 and 8 are instructive in this regard. They are used to illustrate the interference pattern on a spectral range much larger than the experiment. In Fig. 7a the recursion trails of the P-branch of transitions from v'''= 34 is shown. In Fig. 7b, both P- and R- branches of v''' = 34 are included to show the interference between rotational branches of a single vibrational band. Fig. 8 illustrates the third-order polarization for v''' = 26-45 in the B state. The image in Fig. 8a is obtained by summing over the squared transition probabilities, as opposed the squared sum required by Eq. 17 and shown in Fig. 8b. This allows the illustration of the ro-vibrational revivals, with and without interference among the transitions. Remarkably, the intricate, predictable pattern that arises after a few rotational revivals (the slow pattern of interferences in 8b reflects the variation of vibration dependent rotational constants), is expected to last until decoherence, therefore until $t \sim 10^{-9}$s.[57]

### B) First- and second-order coherences – TIFRCARS:

Time-integrated frequency-resolved CARS (TIFRCARS) spectra, as a function of time delay between P and coincident S+P pulses are shown in Figure 9a; with representative temporal and spectral slices shown in Figures 9b and 9c. The data were acquired with non-transform limited pulses, with pump centered at $\lambda_P = 553$ nm ($\Delta\lambda_P = 15$ nm, $\Delta\tau_P = 70$ fs) and Stokes centered at $\lambda_S = 577$ nm ($\Delta\lambda_S = 20$ nm, $\Delta\tau_S = 90$ fs). The time-integrated spectra in Fig. 9c identify the ro-vibrational composition of the prepared third-order coherence. Vibrational recursions which form the high frequency modulation of the time slices in Fig. 9b, appear as stripes in the time-frequency image in Fig. 9a. At *t = 0*, when the three input pulses overlap, a dense, nearly structureless spectrum which arises from multiple interaction diagrams is observed, see Fig. 9b. Upon introducing delays longer than a vibrational period in the excited or ground electronic state, the scattering signal intensity drops, and significantly simplified spectra emerge. We will be concerned with the signal at finite time delays, when the P pulse is delayed or advanced relative to the coincident P+S pair, as defined by Eq. 14. In Figure 10, we show the same data as in Fig. 9, now stretched to time delays of *t = ±25 ps*. The time step in these scans is correspondingly coarser, *Δt = 100 fs*, suitable to obtain images of rotational recursions.

*Evolution in first order coherence is followed over the $t_{12}$ interval, or at negative time according to our convention (14a).* At $\omega_P = 553$ *nm*, the P pulse prepares a packet centered near *v' = 22* in the B state, and the observed period of 370 fs is consistent with the 87cm$^{-1}$ spacing of vibrational levels at this energy: $\tau = \hbar/[E(v=22)-E(v=21)] = 380$ fs. At *0 > t > -2ps*, the waveform is deeply modulated; it consists of sharp vibrational recurrences which lose intensity with time, as the rotational density starts to evolve. Rotational revivals occur with a wide dispersion in time, for *t < -2ps*, leading to the slow rolling background. Once rotational phases spread, vibrational recurrences appear with greatly reduced depth of modulation. As ro-vibrational states which have executed different numbers of periods (windings) overlap, the vibrational recursions blur, and the fast modulation is nearly eliminated at *t< -6 ps*, see Fig. 9b. The spectral slices provide complementary information and check the consistency of interpretation. The spectrum *at t = - 2ps* consists of the vibronic progression B(*v' = 27-34*) → X(*v = 0*). At this early time in the evolution of rotational phases, the band profile of each vibration approximates the thermal distribution of rotational states, with population peaking near *j = 50*. This spectrum also allows to establish that the contribution of transitions terminating on *v = 1* is negligible (less than 5%). As time progresses, see the slice at *t = - 4ps*, the rotational band profiles sharpen and shift to lower *j* states, to states which have not yet evolved out of phase. At later time, at *t = -6ps* and *t = -8ps*, the spectra lose definition, retaining little resemblance to the trivially recognizable spectrum at *t = -2 ps*. The spectral features now consist of a congested set of ro-vibrational resonances, due to overlapping ro-vibrational transitions that fall within the spectral bandpass of the monochromator.

*Evolution in the second-order coherence is followed over the $t_{32}$ time interval, or at positive time according to our convention (14b).* This tracks the evolution of the Raman packet on the X state. The persistent shallow modulation observed in the waveforms in Fig. 9b, occur with a period of 160 fs. This is consistent with a packet near X(v" = 4), which could also be inferred from the Raman resonance condition $\omega = \omega_P - \omega_S$. In contrast, the set of prominent peaks at *0 < t < 1ps* in the 541 nm waveform show a period of 350fs. Quite clearly, these vibrational resonances do not belong in the ground electronic state – they are not part of the Raman packet created by the P+S pulses. Evidently, an additional scattering channel contributes to the signal. This is re-enforced by the congested spectra at positive time (see Fig. 9c), which do not show a simple vibrational progression even at early time. This additional scattering channel cannot be reconciled with the two-electronic state Hamiltonian that we have assumed until now. Resonant scattering over yet another electronic state is implicated. To be certain, we consider the explicit wavepacket simulation of the 2-D image of vibronic CARS within the two-electronic state Hamiltonian.

### C) Numerical simulation of vibronic TIFRCARS:

Ignoring all rotational contributions, the vibronic CARS spectra are simulated by the wavepacket propagation method of Kosloff and Kosloff.[58] The simulations were performed assuming that the laser fields only couple the X and B electronic potentials, which are both described as Morse functions.[59] The various order wavepackets are obtained sequentially according to Eq. 4, by integrating the time-dependent Hamiltonian, starting with:

$$i\hbar \frac{\partial}{\partial t} |\varphi_X^{(0)}(t)\rangle = H_X |\varphi_X^{(0)}(t)\rangle \qquad (19)$$

and evaluating all contributions to the third-order polarization:

$$i\hbar \frac{\partial}{\partial t} |\varphi_\alpha^{(n)}(t)\rangle = H_\alpha |\varphi_\alpha^{(n)}(t)\rangle + \mu_{\alpha\beta} E_l(t) e^{\pm i\omega_l t} |\varphi_\beta^{(n-1)}(t)\rangle \qquad (20)$$

where l = P, S, P' and for n even, $\alpha$ = X and $\beta$ = B while for n odd, $\alpha$ = B and $\beta$ = X. The third-order polarization in this perturbation treatment is represented by the sequential absorption and emission of photons ($-\omega_p$ and $+\omega_p$ respectively, when acting on the ket state). For the laser envelope, $E_l(t)$, a Gaussian of fwhm = 70 fs is used, with a transform limited spectral width of 210 cm$^{-1}$. Taking $\omega_P = \omega_{P'}$ = 550 nm, and $\omega_S$ = 571.5 nm, all permutations of the fields in (20) generate ten different-order packets, which are simultaneously propagated for 20 ps. The time dependent spectra are then evaluated by Fourier transformation of the third-order time dependent polarization along the AS direction:

$$S(\omega, \tau) = \left| \int_{-\infty}^{\infty} dt\, e^{-i\omega t} P_{2k_P - k_S}^{(3)}(t, \tau) \right|^2 \qquad (21)$$

The calculated time-frequency image is shown in Figure 11. The simulated spectra have been convoluted with a Gaussian of fwhm = 70 fs to compensate for the absence of rotations in the comparisons with experiment.

Since, all possible timing diagrams are explicitly included in the computation, their contribution to the final observables can be directly assessed. Thus, at negative time, the AS polarization formally consists of:

$$P^{(3)}_{2k_P-k_S}(t) = \langle \varphi_X^{(0)} | \hat{\mu} | \varphi_B^{(3)}(\omega_{P'}, -\omega_S, \omega_P, t) \rangle + \langle \varphi_X^{(2)}(\omega_S, -\omega_{P'}, t) | \hat{\mu} | \varphi_B^{(1)}(\omega_P, t) \rangle + c.c. \quad (22)$$

while at positive time (see diagrams in Fig. 2):

$$P^{(3)}_{2k_P-k_S}(t) = \langle \varphi_X^{(0)} | \hat{\mu} | \varphi_B^{(3)}(\omega_{P'}, -\omega_S, \omega_P, t) \rangle$$
$$+ \langle \varphi_X^{(2)}(\omega_S, -\omega_{P'}, t) | \hat{\mu} | \varphi_B^{(1)}(\omega_P, t) \rangle + \langle \varphi_X^{(2)}(\omega_S, -\omega_P, t) | \hat{\mu} | \varphi_B^{(1)}(\omega_{P'}, t) \rangle + c.c. \quad (23)$$

For the chosen input pulses, the contribution from the first term in both (22) and (23) is three orders of magnitude larger than the rest of the terms. Hence the importance of the time-circuit diagram in Fig. 1 which has been used in all of our discussions. The closed nature of the time circuit diagrams implies that each initial statistical state can be separately evaluated, and co-added to obtain the overall signal. As such, it is verified that although the thermal occupation of B(v=1) is 35%, its contribution to the final spectrum is negligible (<1%). This is the result of the strong selectivity of the nonlinear CARS process, as expressed by the weights of states in Eq. 18.

The agreement between simulation and data over the $t_{12}$ interval (negative time) is quite acceptable, given that rotations are not included in the simulation. The simulation predicts a rather simple vibrational progression. The first recursion occurs with a relatively compact packet, giving a pulse width limited signal. Due to the anharmonicity of the potential, the recurrences in the signal broaden with time and develop a chirp. By the tenth recurrence, near *t = 4 ps*, the packet splits and shows a doubling in the signal, in good agreement with the experimental 2-D image of Fig 9a.

The predicted image at positive time is rather similar to that at negative time. It consists of a simple vibrational progression, now with a recurrence given by the X-state vibrational period. The simulation does not reproduce the congested and strongly blue shifted spectrum observed in the experiment at positive time (see Fig. 9c). Indeed, in the two-electronic state Hamiltonian, if time evolution is ignored under the laser pulses, then the spectral composition of the CARS signal should be symmetric in time. This would be the expectation if we were to note that states in the third-order superposition are determined by energy conservation, *δ[E-(2ω_P-ω_S)]*. Evolution of the molecular Hamiltonian under the laser pulses breaks this symmetry. The simulations show that at negative time, the spectral intensity peaks near 530 nm, near the wavelength predicted by the energy conservation condition (10a). At positive delay, the spectral maximum of the third order polarization shifts down by two vibrational quanta, to 533 nm. The down-shift is the result of chirp generated by the sequential action of two pulses on a fast evolving packet,[60,61] which is a process inherent to resonant Raman preparation. Despite the spectral shift of two vibrational quanta, inspection of the Wigner distribution of third-order packet created after one period of evolution in either first or second order coherence does not reveal any clear difference. The vibrational superposition is essentially unchanged.

The simulated spectral asymmetry in time is in the opposite direction of what is observed in the experiment in Fig. 9c. We have verified experimentally that the spectral envelopes and their time-asymmetry are sensitive to small delays in time overlap of the nominally coincident S+P pulses. Similarly, although not systematically studied, we have verified that spectral envelopes are sensitive to the chirp in the laser pulses. Either the chirp of the lasers must be included in the simulations, or chirp must be eliminated from the pulses to be more quantitative in this comparison. At present, we do not fully understand the bi-modal spectral distribution observed at positive time (see spectrum at $t = 2$ps in Fig. 9). An intriguing possibility is the interference between preparation of Raman packets via the B and B" surfaces (see Fig. 12). Indeed, interference between these two channels has been identified previously in the analysis of the resonant Raman spectra of iodine in rare gas solids.[62]

### D) $P^{(1,2)}(t)$ contribution – Interference between vibronic packets:

Laser chirp cannot explain the vibrational recurrences that appear at 0<t<1ps in the waveforms for λ = 532 nm – 541 nm in Fig. 9b. This set of recurrences appears most prominently in the 541 nm waveform, where as many as four resonances are observed with a period of 350 fs, over a background modulated at twice higher frequency by the vibrational packet on the X state. The observed period is too slow to be assigned to a packet on the X state, and it is distinctly faster than the 380 fs period of the packet prepared by the pump pulse near *v = 22* of the B state. A packet near *v= 19* of the B state would have the observed period. However, energetically, such a packet could not be prepared with the pulses used. Moreover, if prepared, it would not be expected to decay as rapidly as observed. Since the simulations succeed in capturing the negative time image, and since they contain all timing diagrams of the third-order polarization, we may conclude that these anomalous vibronic resonances involve an additional electronic state in the molecular Hamiltonian.

A consistent interpretation of this scattering channel is obtained by considering the interferometric signal between two vibrational packets in the B state, separately prepared by the pump and Stokes pulses near v = 22 and v = 14, and cross-correlated via a transition to a short-lived excited electronic state. The suggested time-circuit diagram for this process is given in Figure 12. It assumes optical coupling between the B state and a higher lying electronic surface that can be reached with the P'-photon. Energetically, a manifold of I*+I and I* + I* repulsive potentials are accessible. We assume in the figure that the transition is B($0_u$)↔I*I*($0_g$), which we have previously analyzed.[63] To be clear, consider the transcription of the time-circuit diagram of Fig. 12 to the explicit perturbation expression for a given spectral component of the AS polarization:

$$P^{(1,2)}(\omega_{As}) = \int_{-\infty}^{\infty} dt \langle \varphi_B^{(1)}(t;-\omega_s) | \hat{\mu}_{CB} e^{i\omega_{As}t} | \varphi_C^{(2)}(t;\omega_p,\omega_{p'}) \rangle$$

$$= \int_{-\infty}^{\infty} dt \int_{-\infty}^{t} dt_3$$

$$\times \langle \varphi_B^{(1)}(t_2;-\omega_s) | e^{iH_B(t-t_2)/\hbar} \hat{\mu}_{BC} e^{i\omega_{As}t} e^{-iH_C(t-t_3)/\hbar} \hat{\mu}_{CB} E_{p'}(t_3) e^{-i\omega_{p'}t_3} e^{-iH_B(t_3-t_2)/\hbar} | \varphi_B^{(1)}(t_2;\omega_p) \rangle$$

(24)

in which the time origin, $t_2$, is taken as that of the coincident arrival of P+S pulses. For a dissociative upper state, since the resonant scattering is in effect instantaneous, we may suppress evolution past $t_3$. Since the duration of the laser is short in comparison to the period of motion, the snapshot limit is appropriate. Then in terms of the dipole dressed wave packets, $\phi$, we have:

$$P^{(1,2)}(t) = \langle \phi_B^{(1)}(t;-\omega_s) | E(\omega_{p'}) \delta[\omega_p - \Delta V_{CB}] | \phi_B^{(1)}(t;\omega_p) \rangle + c.c.$$

$$= \langle \phi_B^{(1)}(t;-\omega_s) | W(q) | \phi_B^{(1)}(t;\omega_p) \rangle + c.c. \qquad (25)$$

$$= \int dp\,dq\,dp'\,dq' \langle \phi_B^{(1)}(t;-\omega_s) | p',q' \rangle \langle p',q' | W(q) | p,q \rangle \langle p,q | \phi_B^{(1)}(t;\omega_p) \rangle + c.c.$$

where *t* is the time delay between the arrival of S+P and P' pulses. Given an assumed form of the upper state potential, Eq. 24 is easily integrated numerically. Analytical evaluation of Eq. 25 is possible by taking a stationary Gaussian for the window function, *W(q),* and Gaussian packets for the P- and S-prepared superpositions. If we were to assume a window delta in space, then (25) reduces to the cross correlation between the S- and P-prepared packets. This correlation in energy representation, yields the strictly temporal evolution of the signal:

$$P^{(1,2)}(t) = \sum_{v'} \langle \phi_B^{(1)}(t;-\omega_s) | v' \rangle \langle v' | \hat{v} \rangle \hat{P} \langle v'' | \sum_{v''} |v''\rangle \langle v'' | \phi_B^{(1)}(t;\omega_p) \rangle + c.c.$$

$$= \mu_{CB}^2 c(v',v'') \left( \sum_{v'} a_{v'} |v'\rangle e^{-i\omega_{v'}t+\alpha} \right) \left( \sum_{v'} b_{v'} \langle v'| e^{+i\omega_{v'}t+\beta} \right) + c.c. \qquad (26)$$

where

$$a_{v'} = |\mu_{BX}|^2 \sum_v e^{-\beta E_v} |\langle v'|v\rangle|^2 E(\omega_p)\delta[\hbar\omega_p - \Delta E_{vv'}]$$

and (27)

$$b_{v'} = |\mu_{BX}|^2 \sum_v e^{-\beta E_v} |\langle v'|v\rangle|^2 E(\omega_S)\delta[\hbar\omega_S - \Delta E_{vv'}]$$

and the projector from v" to v' in (26) is the stimulated Raman process spelled out in (24). Under the assumption that this process does not color the CARS spectrum we may set the coefficients c(v',v") to a constant. Using the experimental spectral profiles of the P- and S-pulses, the vibrational amplitudes in each of the prepared superpositions are defined according to (27), and the expected time-dependent CARS signal obtained according to (26). The simulated $P^{(1,2)}(t)$ contribution to the signal is shown in Fig. 13. Note, the scattering at $t = 0$ has contributions from several diagrams, and therefore will be modulated as a function of AS wavelength. This is illustrated in Fig. 13 by showing the waveforms obtained at two wavelengths. The spectral dispersion of the waveform is given in (24). The observable signal can also be understood as the cross-correlation between the two packets. Thus, for an S-prepared packet centered at v = 14 with a mean vibrational spacing of $\bar{\omega}_S$ = G(v=14) − G(v=13) = 102.6 cm$^{-1}$ (period = 325 fs), and a P-prepared packet centered at v = 22, with $\bar{\omega}_p$ = G(v = 22) − G(v=21) = 87.6 cm$^{-1}$ (period = 380 fs); the signal is obtained by squaring (26):

$$I^{(1,2)}(\omega_{As},t) = I\cos(\frac{\bar{\omega}_S + \bar{\omega}_p}{2}t + \vartheta)\cos(\frac{\bar{\omega}_S - \bar{\omega}_p}{2}t + \vartheta)e^{-t/\tau} \qquad (28)$$

The signal contains three characteristic time constants:
   a) A fast modulation at the center frequency, $\tau_1 \sim 2/(\bar{\omega}_S + \bar{\omega}_p)$ = 350 fs, corresponding to the superposition moving out of the window as a whole;
   b) A slower modulation at the difference frequency, $\tau_2 = 2/(\bar{\omega}_S - \bar{\omega}_p)$ = 2 ps, corresponding to decay of the signal as the S- and P-prepared packets split;
   c) An even slower decay envelope given by the dispersion of frequencies within each distribution, namely, the spreading of the individual packets dictated by the local anharmonicities on the B state, $\tau_3 = 1/\omega_e x_e$ = 15 ps.

The slower decay time, $\tau_3$, is now convoluted with the rotational dephasing which occurs on a similar time scale; and intensities in signal recurrences will now depend on the rotational phase distribution. Note, despite the decay of the modulation, this channel will contribute to the scattering process at all positive times, responsible in part for the broad time-integrated spectra at positive times in Fig. 9c.

Additional electronic resonances can be inferred from the known dense manifold of electronic states of iodine, and from experiments in which we vary the time ordering in the sequence of non-overlapping pulses. As example, note that the B" state (Fig. 3) which is directly accessible from the ground, must contribute to CARS at positive delays. Under coincident P+S pulses, packets created by the P pulse on both the B and B" surfaces may be transferred to the X state with the S pulse. Due to the difference in the curvature of potentials, the B and B" channels of scattering would create a bimodal vibrational distribution in the X state, which would then be reflected in the AS spectrum. However, since the B" state is dissociative, it cannot contribute to CARS at negative delays. A delay of <20 fs after the P pulse would be sufficient to ensure that the B" packet escapes further interrogation.

In short, there are several distinct electronic scattering channels that are observed to contribute to the CARS signal at positive delay, while a single channel dominates the signal at negative delay.

### E) The ro-vibrational contribution to first- and second-order images:

The ro-vibrational contribution to the time-integrated CARS spectrum can be written down with the help of the diagram in Fig. 4. Since the detection involves integration over $t_4$, the signal is determined by the time intervals $t_{32}$ and $t_{21}$. For a given path in vibrational state space, $v',v'',v'''$, three rotational circuits contribute to a given ro-vibronic transition in the CARS spectrum. Thus, for a P($j$) transition we have:

$$I^{(3)}_{P(j)}(v',v'',v''') = \{(2j+1)e^{\beta E_{v,j}}[c_1 e^{-i[\omega_{v',j+1}t_{21} + \omega_{v'',j}t_{32} - \omega_{v,j}t_{31}]}$$
$$+ c_2 e^{-i[\omega_{v',j-1}t_{21} + \omega_{v'',j}t_{32} - \omega_{v,j}t_{31}]} \quad (29)$$
$$+ c_3 e^{-i[\omega_{v',j-1}t_{21} + \omega_{v'',j-2}t_{32} - \omega_{v,j}t_{31}]} + c.c.]\}^2$$

in which the weighting coefficients are the transition matrix elements encountered in (18). Ignoring the slowly varying coefficients, namely rotational matrix elements and F-C factors, but retaining the energy conservation condition derived from the spectral composition of the pulses $c(v',v'',v''') = \delta[E_L - E_{v,j}]^3$, it is possible to reproduce the experimental waveforms of Figure 8 and 9. At positive delay ($t_{32} = 0$), the AS spectral line intensity will be modulated according to:

$$I^{(3)}_{P(j)}(v',v'',v''',t_{21}) = \{(2j+1)e^{\beta E_{v,j}} \sum_{v',v''} c(v',v'',v''')[e^{-i(\omega_{v',j+1} - \omega_{v,j})t_{21}} + 2e^{-i(\omega_{v',j-1} - \omega_{v,j})t_{21}} + c.c.]\}^2$$

(30a)

while at negative time ($t_{21} = 0$):

$$I^{(3)}_{P(j)}(v',v'',v''',t_{32}) = \left\{(2j+1)e^{\beta E_{v,j}}\sum_{v',v''}c(v',v'',v''')[2e^{-i(\omega_{v'',j}-\omega_{v,j})t_{32}} + e^{-i(\omega_{v'',j-2}-\omega_{v,j})t_{32}}+c.c.]\right\}^2$$

(30b)

These impulse response functions, when convoluted with the laser cross-correlation function yield the finite pulse signals. Carrying out the square in (30) yields the rotational recursions at beat frequencies:

$$[(E_{v',j+1} - E_{v,j}) - (E_{v',j-1} - E_{v,j})]/\hbar = 4\overline{B}(v')j \qquad (31a)$$
$$[(E_{v'',j} - E_{v,j}) - (E_{v'',j-2} - E_{v,j})]/\hbar = 4\overline{B}(v'')j \qquad (31b)$$

where B(v') and B(v") are the vibration dependent rotational constants in the excited and ground electronic states, respectively; and the over-bar implies averaging over the v',v" states that are accessible under the broadband laser pulses.

In Fig 14 we provide a comparison between simulation (30) and an experimental waveform sliced from the two-dimensional image in Fig. 9. The simulation uses (30a) and (30b) separately, and joins them at *t = 0*. This results in the mismatch of amplitude for the peak at *t= 0*. At *t= 0*, since $t_{21} = t_{32}$, the two diagrams shown as insets are degenerate, and therefore interfere (destructively). On this time-scale since the evolution is entirely vibrational, the effect is reproduced in the explicit wavepacket simulations of Fig. 11. To match the depth of modulation in experiment and simulation, we have used a convolution width of 70 fs in $t_{12}$, and a width of 140 fs in $t_{23}$. This effective reduction of modulation depth for interrogating the Raman packet is expected, and discussed in some depth previously.[16] It is a result of the separation between positive and negative momentum components of the Raman packet as it enters the resonance window.

In Fig. 15 we provide the same comparison as in Fig. 14, but now for the coarse grain, long-time scan of Fig. 10a. Although, single rotational lines are resolvable, the experimental spectra are instrument limited by the spectrometer bandpass of 8 cm$^{-1}$. Thus in the waveform of Fig. 14 which corresponds to the spectral slice at 538 nm, the main contributing transitions are: R54(29-0), P50(29-0), R89(30-0), P85(30-0), R111(31-0), P108(31-0) where in parenthesis we give the vibrational transition. Note, in contrast with the third-order coherence obtained with gated detection, in this case the lines that overlap in the observation window do not interfere - they are simply convoluted with the spectral bandpass. The use of short pulses forces the fan-out in vibrational state space. In the present v' = 23-28 and v" = 2-5 are the main contributors to the signal. The vibration dependence of rotational constants (due to coriolis, centrifugal and higher order distortions) generates dispersion in rotational recurrences even when a single line is being monitored. Accordingly, the rotational recursions of Fig. 10 can be understood, as resulting from the overlap of several different ro-vibrational lines in the spectral bandpass of detection. The prominent recursion is that of the thermal maximum in the rotational population, *j~50*, for which the $\overline{B}(v') \approx B(v'=25) = 0.02414$ cm$^{-1}$,[55] which leads to a period of 6.9 ps (31a) as seen in Fig. 15 (we have not included higher order corrections to the rotational constant in the simulations). For the Raman packet of Fig. 10b, the dominant recursion occurs at t ~ 4.5 ps, consistent with $\overline{B}(v'') \approx B(v''=3) = 0.036966$ cm$^{-1}$.[55] Due to approximations made in weighting coefficients and spectroscopic constants, the match between experiment and simulation in Figures 14 and 15 is not exact. Nevertheless, the comparison is sufficiently detailed to confirm our analysis of the manipulated molecular coherences.

## IV. DISCUSSION

The detailed exposition of the experimental time-frequency resolved CARS measurements and their analysis in vapor iodine serves mainly for the purposes of understanding the four-wave mixing process with the molecule acting as mixer. Despite the fact that we are interrogating a diatomic, the participation of a large number of rotational states in the molecular coherence makes the analysis valuable. The time-frequency images are understood and reproduced in terms of phase evolution in multi-dimensional state space; formally, in the Hilbert space, H, consisting of the tensor product of electronic, vibrational, and rotational spaces: $H = H_{el} \otimes H_{vib} \otimes H_{rot}$. The time-circuit diagrams and the phase coherence condition (8) provide the necessary bookkeeping to describe the various order coherences, as demonstrated by reconstructing the experimental two-dimensional images. The images contain information. The interferometric nature of the images, most clearly illustrated for the third-order coherence (figures 6-8), implies that the encoded information retains the wave nature inherent in quantum amplitudes. Such data can be used to reconstruct the molecular Hamiltonian, by extracting accurate spectroscopic constants. Also, the significant control exercised over the evolving molecular coherence through the three input laser fields can be taken as a paradigm for molecular control. The latter aspect, for the important case of single-color four-wave mixing has been recently explored.[61,64] A useful application of quantum control is to be expected in quantum computation, or quantum information transfer, which we explore here.

The observable coherent polarization is the outcome of operations on a quantum register consisting of the superposition of product states $|\rangle_{el}|\rangle_{vib}|\rangle_{rot}$. The nontrivial superposition of product states provides entanglement,[65] an aspect unique to quantum information and key to scalable parallelism in quantum computing.[9,10] In the present case, the Born-Oppenheimer separated molecular Hamiltonian offers the ro-vibronic Hilbert space of $2 \times m \times n$ (el$\otimes$vib$\otimes$rot) dimensions with the possibility of three natural entanglements. While the dimensionality is quite large, m~10 and n~10$^2$, the structure of this space does not conform to the theoretically optimal structure of $2^N$ dimensional space, namely, the space of N qubits. Nevertheless, given efficient parallel logical operations on a register of 10$^3$ elements and a processor speed of 10$^{-15}$s - 10$^{-12}$s, significant applications can be expected. Tasks that rely on few entanglements include Grover's search algorithm,[20] and quantum cryptography.[66] The mundane building blocks for such applications – reset, logical operation, and readout – can be readily inferred from the data presented. In what follows we first consider the measurement and readout process more closely, then present the minimal but sufficient set of efficiently executable universal logic gates for all-purpose quantum computing. It should be clear from the onset that the stored and retrieved information is in the complex amplitudes of eigenstates that define the evolving coherences. Process control is provided by the coherence of the laser fields, which can be thought to consist of a time sequence of discrete spectral components.

### A) Measurement and Readout

There is a fundamental difference between time-gated detection and time-integrated detection of the anti-Stokes polarization. The first allows readout by projection on superposition states, while the latter projects out a single component of the evolving polarization for observation. This is directly illustrated for the electronic-rotational entanglement as discerned from the rotational recurrences in various order coherences.

When using time-gated detection, the prominent rotational recurrence frequency in third-order coherence occurs at $|2(B'-B")j|$, at the beat between consecutive P-branch or R-branch transitions of the AS polarization (B' and B" are the rotational constant of the molecule in the B and X electronic state, respectively). Noting that the classical rotation frequency of a diatomic is *2Bj*, the observed recurrence is recognized as the difference between classical periods of rotation on the ground and excited electronic surfaces. It would appear that over the $t_{43}$ interval the molecule rotates forward on one electronic surface and backward on the other. The observable period of rotational recurrence reports a property of an electronically maximally entangled state: $|\varphi\rangle = \sum a|X\rangle|v\rangle|j\rangle + \sum b|B\rangle|v'''\rangle|j'''\rangle$, with $\left(\sum a\right)^2 = \left(\sum b\right)^2 = 0.5$.[67]

In contrast, when using time-integrated detection, the rotational revivals occur (31) with a frequency of *4B'j* in first-order coherence, during $t_{21}$, and at *4B"j* in second-order coherence, during $t_{32}$, *i.e.,* at twice the classical frequency of rotation in the excited and ground state, respectively. Despite the fact that in first-order the evolution is in an electronic coherence while in second-order it is a vibrational coherence that evolves as an electronic population on the ground electronic surface, in both cases the property of a single electronic state is measured. It would appear that the molecule is rotating in the excited electronic state during $t_{21}$, just as it must rotate in the ground electronic state during $t_{32}$. These results can be understood using the time-circuit diagram of Fig. 4b. With $t_4$ integrated out, and either $t_{12} = 0$ or $t_{23} = 0$, only two circuits contribute to a given transition. In the observable beat between two such circuits, the common path over $\langle v,j|$ is cancelled (31), yielding the beat between states separated by *2j'* or *2j"*. While this explains the signal, more fundamental is the recognition that although over $t_{21}$ the system evolves in an electronic superposition, Fourier filtering over $t_4$ projects out the excited electronic state, as recognized by observing a rotational frequency that reflects the rotational constant of the B state.

TGFCARS detection projects out all transitions that fall under the time-frequency bandwidth of detection. A dramatic manifestation of this is the period of silence at *2 < $t_4$ < 10 ps* in the third-order image of Fig. 6. This is the time interval in which complete destructive interference occurs among the sampled transitions. In terms of the Bloch sphere, by virtue of the equal electronic superposition $(|B\rangle+|X\rangle)/\sqrt{2}$ the system evolves in the $\pi/2$ plane, and the silence occurs as the rotational phases span a flat distribution over all azimuthal angles. This occurs despite the fact that a thermal, statistical density serves as initial state, and therefore the radiators in different *j* states are not initially correlated. The sudden preparation of the initial state, and the fact that the CARS signal is due to bulk polarization, establish a well defined phase among the independent molecules. An even clearer demonstration of inter-molecular coherence in CARS is the observation of beating between vibrations of different molecules in liquid solutions.[68]

The above considerations of rotation-electronic entanglement apply equally well to vibronic coherences. For example, in time-integrated detection, the excited vibrational period is observed during $t_{21}$ despite the fact that the system is in an electronic superposition. In time-gated detection, a vibrational beat between excited and ground state packets occurs, which is effectively smoothed out in the experiments due to the width of the Kerr-gate.

A corollary to the observation that all transitions that fall under the time-frequency gate of detection interfere, is that any selected pair (or sets) of transitions can be made to interfere as long as they can be brought under a given time-frequency gate. Such "hard-wiring" can be accomplished experimentally, *e.g.,* by dispersing the AS polarization then recombining selected spectral ranges on a single detector element after a suitable delay line. Alternatively, instead of using an impulse to drive the time gate, it is possible to use a temporally modulated gate field to detect pre-selected superpositions, *i.e.,* heterodyne detection of CARS using a specifically tailored local oscillator. Such hardwiring creates selective quantum correlations, in effect, producing entanglement through measurement.

### B) Universal Logic Gates

The universal logic gates sufficient to demonstrate a quantum computer consist of the one-qubit operations and the two-qubit controlled-not (*CNOT*) gate.[69] The full set of one-qubit operations is given by the Pauli spin matrices.[70] Here, they can be readily demonstrated on the rotational coherences, and the vibration-electronic entanglement is a readily available two-qubit *CNOT* gate. We first offer the more suggestive representation of the single qubit operations in terms of classical logic gates.

### i) One qubit Operations

Consider the time integrated signal, and in particular, of the spectrally resolved $j\text{-}1 \rightarrow j$ transition illustrated in Fig. 4a. Let us assign the logical value "1" to the presence of this line, and "0" to its absence in the TIFRCARS spectrum. The upper level of this transition $|B,v''',j\text{-}1\rangle$ is connected to the two coherent inputs $e^{i\Omega(v'',j)}|X,v'',j\rangle$ and $e^{i\Omega(v'',j-2)}|X,v'',j-2\rangle$, with phases $\Omega(v'',j;t_2)$ and $\Omega(v'',j) = [\omega(v'',j) - \omega(v,j)]t_{32} + \Omega(v'',j;t_2)$ determined relative to the freely evolving bra-state (see Fig. 1). As in the case of the simulations in Figures 14 and 15, here too, we neglect differences in amplitude due to transition matrix elements. Then for the experimental data where $t_{21} = 0$, the relative phase between the two input lines is determined by the time interval $t_{32}$, namely the time delay between S- and P'-pulses. Assigning the logical values "1" for $\Omega = 2n\pi$ and "0" to $\Omega = (2n\text{-}1)\pi$ for integer n, we may construct a truth table for the visibility of this particular spectral component:

$$\begin{array}{r} 1100 \\ \oplus\ \underline{1010} \\ 1001 \end{array} \quad (33)$$

This bivalent representation of interference between two input channels can be recognized as the inverted exclusive-OR gate, $\overline{XOR}$, (*XOR*, had we assigned "1" to the absence of the line from the spectrum). The *XOR* gate allows modulo 2 addition, which is described by the $\oplus$ operation. The same consideration applied to all nodes in Figure 4b produces the equivalent circuit of Figure 16. The figure describes the bivalent logic of the coherence transferred around the time-circuit diagram of a single initial ro-vibrational eigenstate. If we were to consider the coherent superposition prepared by the P-pulse at $t = t_1$, as the input qubit:

$$|\ \rangle_{in} \equiv a'|B,v',j+1;t_1\rangle + b'|B,v',j-1;t_1\rangle \quad (34a)$$

and identify the output qubit as the coherent superposition prepared at $t = t_3$, after action of the S- and P'-pulses:

$$|\ \rangle_{out} \equiv a'''|B,v''',j+1;t_3\rangle + b'''|B,v''',j-1;t_3\rangle \quad (34b)$$

then using the diagram of Fig. 16 it is easily shown that if fully connected, the circuit flips the logical states between input and output ($a' \rightarrow b'''$ and $b' \rightarrow a'''$). Indeed, this is the bivalent prescription for the simulation of the TIFRCARS signal for the second order coherence shown in Fig. 15 and given by (30) for the continuous representation.

The circuit of Fig. 4b is more flexible than what is inferred by the bivalent logic. The relative phase in the superpositions evolve continuously. Moreover, in TGFCARS detection the relative phase in the output qubit is accessible information. To keep track of this information, we re-write Eq. 29 in the two-dimensional space of the qubit.

$$|\rangle_{out} = \hat{\mu} E_{P'}(t_3) U(t_{32}) \hat{\mu} E_S(t_2) U(t_{21}) |\rangle_{in}$$

$$= \tilde{U}(t_{32}) \tilde{U}(t_{21}) \begin{pmatrix} a' \\ b' \end{pmatrix} = \tilde{U}(t_{31}) \begin{pmatrix} a' \\ b' \end{pmatrix} \quad (35)$$

In the snapshot limit, since the ro-vibrational phases are frozen during the vertical transitions, the entire evolution reduces to one of phases over paths prescribed by the timing of the spectrally broad, temporally sharp pulses:

$$\tilde{U}(t_{21}) = \begin{bmatrix} e^{i\Omega(v',j+1)} & \\ & e^{i\Omega(v',j-1)} \end{bmatrix} = e^{i\Omega} \begin{bmatrix} 1 & \\ & e^{i2B'(2j+1)t_{21}/\hbar} \end{bmatrix} \quad (36)$$

with overall phase $\Omega = (T_e + E_{v'} - E_v + E_{j+1}^{B,v'} - E_j^{X,v}) t_{12}/\hbar$;

$$\tilde{U}(t_{32}) = e^{i\chi} \begin{bmatrix} e^{i\chi_2} + e^{i\chi_0} & e^{i\chi_0} \\ e^{i\chi_0} & e^{i\chi_{-2}} + e^{i\chi_0} \end{bmatrix} = e^{i\chi} e^{i\chi_0} [1 + \sigma_x] e^{i\chi_2} \begin{bmatrix} 1 & \\ & e^{i(\chi_{-2} - \chi_2)} \end{bmatrix} \quad (37)$$

with overall vibronic phase $\chi = (T_e + E_{v''} - E_v) t_{23}/\hbar$; and $\chi_i = (E_{j+i}^{X,v''} - E_j^{X,v}) t_{23}/\hbar$. All rotations under SU(2) are accessible under the timing conditions already realized in the presented data. Thus,

for $t_{32} = 0$ and $t_{21} = (2n-1)\pi\hbar/[2B'(2j+1)]$

$$\tilde{U}(t_{31}) = \tilde{U}(t_{21}) = e^{i\Omega} \begin{bmatrix} 1 & \\ & -1 \end{bmatrix} = 2e^{i\Omega} \sigma_z \quad (38)$$

This simply states that the free evolution of the qubit in the $t_{21}$ interval corresponds to rotation around the z-axis on the Bloch sphere. This rotation occurs with *4B'j* periodicity (38a), namely, the recursion period of the first order coherence (31a). If we do not take advantage of the overall vibronic phase, then to execute $\sigma_z$ rotation on the manifold of *j* states, it would be necessary to use a chirped pulse, with a coherence similar to the third order polarization of Fig. 6. The fast vibronic phase, $\Omega$, allows the execution of the transformation on all *j* states in parallel, by using a short pulse.

For $t_{21}=0$, $\chi_0 = 2n\pi$, $\chi_2 - \chi_{-2} = 2n\pi$; therefore $t_{32} = n\pi\hbar/[2B'(2j-1)]$ and $\chi + \chi_2 = (2n-1)\pi$:

$$\tilde{U}(t_{31}) = \tilde{U}(t_{32}) = \begin{bmatrix} 0 & 1 \\ 1 & 0 \end{bmatrix} = 2\sigma_x \qquad (39)$$

Since $\chi_0 = (E_j^{X,v''} - E_j^{X,v})t_{23}/\hbar = (B^{X,v''} - B^{X,v})t_{23}/\hbar$, and rotational constants of different vibrations within the same electronic state are similar, $\chi_0 \sim 0$ is easily fulfilled. As before, it is necessary to take advantage of the vibronic overall phase, $\chi$, for parallel execution of $\sigma_x$ rotation on the manifold of rotational qubits.

The execution of a $\sigma_y$ rotation requires finite $t_{21}$ and $t_{32}$. The required conditions are readily obtained. In principle this is superfluous, since arbitrary rotation around two independent axes is sufficient to span the full space of the two-dimensional qubit.

ii) Two-qubit controlled gate

A useful two-qubit CNOT gate would be one that operates on entangled states, using qubits in different spaces as target and control.[71] Until now, the electronic and vibrational qubits have been used as simple state tags, with their states determined by the order of interaction with the laser field. This determinism in the electronic state is the result of choosing P. S. P' colors that yield only one diagram in the four-wave mixing process. To control electronic qubits arbitrarily, it is necessary to create a multiplicity of interfering paths. This is realized in single-color four-wave mixing, or for the condition where all three input fields overlap spectrally, since under such conditions CARS and CSRS paths can interfere. As an illustrative example, consider the vibrational phase of an input vibronic qubit to control the electronic superposition in the output qubit.

At the expense of greatly reducing the vibrational vector space, let us return to the wavepacket picture $|\varphi_v\rangle$ and assign it a logical value $|1\rangle$ when the packet is in the Frank-Condon window of the X↔B transition, and $|0\rangle$ otherwise. To further conform to quantum logic implementations, let us code the electronic qubit as $|0\rangle \equiv |X\rangle$ and $|1\rangle \equiv |B\rangle$. Then, after preparation of the initial coherence with the first pulse, the action of the pulse at $t=t_2$ can be defined as:

$$|\varphi_v\rangle|\varphi_{el}\rangle \xrightarrow{E_S(t_2)} |\varphi_v\rangle(|\varphi_v\rangle \oplus |\varphi_{el}\rangle) \qquad (40a)$$

Which reads: if the vibrational packet is in the F-C window, then change the electronic state. This can be verified to be the two-qubit *CNOT* gate with the logical values

$$|1\rangle|0\rangle \longrightarrow |1\rangle|1\rangle$$
$$|1\rangle|1\rangle \longrightarrow |1\rangle|0\rangle$$
$$|0\rangle|0\rangle \longrightarrow |0\rangle|0\rangle \quad (40b)$$
$$|0\rangle|1\rangle \longrightarrow |0\rangle|1\rangle$$

Depending on $t_2$, the operation in (40) controls the electronic superposition that will evolve in the $t_{32}$ interval. Thus, taking $\omega_X$ and $\omega_B$ as the frequencies of vibrational packets in the X and B states, respectively; starting with the coherence $|\varphi_{v'}\rangle|B\rangle\langle X|\langle\varphi_v|$ prepared by the pump pulse; the possible outcomes of the action of the pulse at $t_2$ are:

$$|\varphi_{v'}\rangle|B\rangle\langle X|\langle\varphi_v| \xrightarrow{E(t_2)} |\varphi_{v'}\rangle|B\rangle\langle B|\langle\varphi_v| \quad \text{if } t_{21} = 2n\pi/w_X$$
$$\xrightarrow{E(t_2)} |\varphi_{v'}\rangle|X\rangle\langle X|\langle\varphi_v| \quad \text{if } t_{21} = 2n\pi/w_B$$
$$\xrightarrow{E(t_2)} a|\varphi_{v'}\rangle|X\rangle\langle X|\langle\varphi_v| + b|\varphi_{v'}\rangle|B\rangle\langle B|\langle\varphi_v| \quad \text{if } t_{21} = 2n\pi/w_B = 2m\pi/w_X$$
(41)

When $t_{21} = 2n\pi/\omega_X$, the bra-packet on the X electronic surface is in the F-C window when the second pulse arrives, therefore an electronic population is created in the B state: $|0\rangle|B\rangle\langle X|\langle 1| \rightarrow |\varphi_{v'}\rangle|B\rangle\langle B|\langle\varphi_v|$. When $t_{21} = 2n\pi/\omega_B$, the ket-packet on the B electronic state is in the F-C window when the second pulse arrives, therefore an electronic population is created on the X surface: $|1\rangle|B\rangle\langle X|\langle 0| \rightarrow |\varphi_{v'}\rangle|X\rangle\langle X|\langle\varphi_v|$. Finally, when $t_{21} = 2n\pi/\omega_B = 2m\pi/\omega_X$, when the packets on the B and X surfaces are simultaneously in the F-C window, an electronic superposition is created: $|1\rangle|B\rangle\langle X|\langle 1| \rightarrow a|\varphi_{v'}\rangle|X\rangle\langle X|\langle\varphi_v| + b|\varphi_{v'}\rangle|B\rangle\langle B|\langle\varphi_v|$. Thus, at a given delay, the electronic superposition to be prepared is controlled by the phase of the vibrational packet. In effect, in single-color four-wave mixing, the contributions from various diagrams can be controlled, $P^{(3)}=aP^{(0,3)}+bP^{(1,2)}$, to drive the system to the desired state. Indeed, these timing conditions have been experimentally verified in degenerate four-wave mixing studies on $I_2$.[37,37,61,64]

Similar controlled gates are possible between rotation-vibration, and rotation-electronic qubits.[72] Here, we are satisfied by indicating that the requisite two-qubit *CNOT* gate and the one-qubit operations are naturally wired in molecular four-wave mixing, sufficient to identify the potential of the system for quantum computing. A schematic of the general conceptual approach to using the present network for a single path in vibrational vector space is provided in Fig. 17. In the example, a broad-band P-pulse resets the quantum register, a structured S-pulse writes, the P'-pulse is used to process, and the output register is read either in TIFRCARS mode, or in TGFCARS with heterodyne detection using the gate pulse (G-pulse) as local oscillator.

## V. CONCLUSIONS

TFRCARS yields multi-dimensional images of molecular coherences, which we have presented here as a set of three two-dimensional plots. The images of the various order molecular ro-vibrational coherences contain sufficient detail to allow an accurate characterization of the molecular Hamiltonian. Although it has been previously demonstrated that time resolved measurements can provide spectroscopic constants of high accuracy,[73] and the present method of extracting two-dimensional images has the advantage of multiplexing, such an application to stable small molecules is of limited value. The unique advantage of the method as a spectroscopic tool is in its ability to characterize transient spectra, or transient species. This derives from the fact that the method permits transform limited observation and interrogation of the conjugate time and frequency coordinates. Consider the application of the method to image chemistry. The simplest example would be that of unimolecular dissociation. If the process is slower than the speed of the Kerr shutter in use, as in rotational predissociation,[74] then it should be possible to directly record the ro-vibrational tracks of the dissociating complex in third order-coherence using TGFCARS. An image, similar to that in Fig. 6 would now provide a map of the ro-vibrational channels to bond-breaking. More generally, consider evolution along a reactive coordinate $Q(t)$ that has been set in motion via a short optical pulse. Then the second-order coherence would image the evolution of frequencies orthogonal to $Q$, the stiffening or loosening of bonds due to chemistry (or any other change). If the pulses used are shorter than the time scale of evolution along $Q$, direct images are obtained. Otherwise, a deconvolution similar to that used in the frequency resolved optical gating (FROG) to characterize laser pulses,[11] would be applied. Indeed, the time gated photon echo measurements in the liquid phase accomplish these very aims, with evolution along the solvation coordinate being the target of interest.[39-42] The same principles applied to protein unfolding and excitonic dynamics have been given recently.[75,76]

TFRCARS, or more generally four-wave mixing experiments, can be used as a method for molecular coherence control. This concept is usually associated with the possibilities of controlled chemistry. In practice, such applications are likely to be somewhat limited.

Time-frequency-resolved four-wave mixing using a molecule as mixer, allows the preparation, manipulation, and readout of massive quantum superpositions with sufficient control to consider applications to quantum computing. Parallels can be drawn between this approach and NMR,[77] which is the maturer field of coherence control.[78] For example, the $P^{(0,3)}$ polarization can be regarded as stimulated photon reverberations, in analogy with the NMR photon echo.[79] The optical four-wave mixing approach has important advantages with regard to the manipulation and transfer of massive coherences, which is an important step toward practice in quantum computation. Since the third-order polarization is the observable, the optical process does not require polarization of the thermal initial ensemble, which is required in NMR and a major source of the limitation to few qubits.[80] The optical four-wave mixing method circumvents this, since it allows the sequential manipulation of the initial coherence struck by the first pulse on a thermal background that does not interfere with the signal. Moreover, since the manipulations involve optical pulses, single operations can be accomplished on fs-ps time scales. Given the Doppler control of decoherence in rarified media at room temperature, $10^3 - 10^4$ operations can be completed prior to loss of signal. Clearly a major challenge in the proposed approach is in cascading, to enable multiple sequential operations. Note, at least in oracle type applications single qubit logic gates can be implemented with efficiency and parallelism, and the required quantum *CNOT* gate can be readily implemented in one step. Recognizing that the observable output in the four-wave mixing experiment is a coherent radiation field, it is not too difficult to imagine sequential processing, or networking, by having the AS beam from one stage act as one of the input fields for a second stage. The considerations we have presented, we believe, are sufficient to encourage a search for useful algorithms that naturally map on molecular networks.

## VI. ACKNOWLEDGMENTS

The support of this research through grants from the US AFOSR (F49620-98-1-0163) and the NSF (9725462), is gratefully acknowledged. Discussions with Z. Bihary and J. Eloranta were most valuable in formulating some of the quantum-computational concepts presented here. Discussions with A. Ouderkirk on physical designs of four-wave mixing cascades to construct logic networks, is fondly acknowledged.

**Figure Captions**

**Fig. 1:** Diagrammatic representation of time-resolved CARS. Both time-circuit and Feynman diagrams are illustrated for a non-overlapping sequence of P, S, P' pulses, with central frequency of the S-pulse chosen to be outside the absorption spectrum of the B←X transition, to ensure that only the $P^{(0,3)}$ component of the third-order polarization is interrogated. In this dominant contribution, all three pulses act on bra (ket) state while the ket (bra) state evolves field free. Note, for the Feynman diagrams, we use the convention of Ref. 5, which is different than that of Ref. 4.

**Fig. 2:** Diagrammatic representation of electronically all-resonant CARS with coincident Pump and Stokes pulses, P+S, followed by P'. The first diagram is the same as that of Fig. 1. To be resonant in all fields, the $\langle \varphi^{(2)} | \hat{\mu} | \varphi^{(1)} \rangle \equiv P^{(2,1)}$ contribution can only be initiated from vibrationally excited states of the ground electronic surface. The time-circuit diagrams make it clear that for this process to have significant cross section, $\omega_p - \omega_s \sim k_B T$ must hold.

**Fig. 3:** The wavepacket picture associated with the evolution of the ket-state in the diagram of Fig. 1, for resonant CARS in iodine. The required energy matching condition for the AS radiation, Eq. 10b of text, can only be met when the packet reaches the inner turning point of the B-surface. Once prepared, $\varphi^{(3)}(t)$ will oscillate, radiating periodically every time it reaches the inner turning point.

**Fig. 4**: The "wiring" diagram of rotational eigenstates for a given path – v, v', v", v''' – in vibrational state space.
    a) Conventional diagram: The rotational selection rule, $\Delta j=\pm 1$, limits the possible paths. The dark solid lines connect the paths that lead to the P-branch ($j-1 \rightarrow j$) transition in the CARS spectrum. These are the only

relevant paths when the measurement involves TIFRCARS of the *j-1→j* transition. The gray lines connect the path for the R-branch transition, *j+1→j*. Interference occurs when two transitions terminate on one eigenstate. Note also that the coherence transferred to *<v''',j±3|* is lost from the detectable AS polarization.

b) Schematic diagram, useful for the interferometric analysis of the four-wave mixing process. The diagram highlights the phases gained by evolution along different paths in ro-vibronic state space. Each inner loop is equivalent to a Mach-Zehnder interferometer, with nodes corresponding to a time at which the applied radiation field may stimulate interference between two paths.

**Fig. 5:** Experimental arrangement for TGFCARS. The sample consists of iodine vapor contained in a quartz cell heated to 50°C. The forward BOXCARS geometry is used, with the three input beams brought to focus using a 25 cm achromatic doublet, and a pinhole to spatially filter the anti-Stokes radiation. The AS beam is *t*-gated using a Kerr cell with $CS_2$ as active medium, then dispersed through a 1/4-m monochromator, and detected using a CCD array. An experimental spectral slice, at t = 2 ps, is shown in the inset.

**Fig. 6:** Image of the third-order ro-vibronic coherence obtained by time gated frequency resolved CARS (TGFCARS): Experiment (right panel), simulation (left panel). The vibrational assignment is indicated at the population maxima, near *j = 50*, for each state. Note, the signal at *t = 0* is strongly saturated.

**Fig. 7**: Simulation of the third-order coherence for v''' = 34 showing the rotational revivals, or winding pattern of rotational recursions: a) P-branch transitions; b) P- and R-branch transitions. Note the interference between branches, and within the same branch.

**Fig. 8**: Simulation of the role of interference in the third-order polarization, for v''' = 26-45. a) The signal evaluated according to Eq. 17 of text, under the assumption of white spectra for the three pulses, and after setting all matrix elements to unity. b) The third order polarization with interferences suppressed by squaring contributions from each eigenstate before summing, as required by Eq. 17.

**Fig. 9**: Time-integrated frequency-resolved CARS image of first- and second-order coherences and cross sections: a) Two-dimensional time-wavelength image; b) Waveforms (time-slices) at selected wavelengths; c) Spectra at selected times. Negative time corresponds to scanning $t_{21}$, with $\delta[t_3-t_2]$, during which the first-order electronic coherence is monitored. Positive time corresponds to scanning $t_{32}$, with $\delta[t_2-t_1]$, during which the second-order vibrational coherence is monitored.

**Fig. 10**: Coarse grain images of time-integrated frequency resolved CARS: a) First-order coherence, where $t = t_{21}$; b) Second-order coherence, where $t = t_{32}$.

**Fig. 11**: Wavepacket simulations of first- and second-order coherence images of Fig. 9a, within the two-electronic state Hamiltonian.

**Fig. 12**: Time circuit diagram for the resonant $P^{(2,1)}$ contribution and the likely potential energy surfaces that are involved.

**Fig. 13:** The $P^{(1,2)}$ contribution simulated, using Eq. 26 of text, based on the experimental spectral profile of the laser pulses, Eq. 27.

**Fig. 14**: Spectral slice of the TIFRCARS signal in Fig. 9b at 538 nm. The peaks identified with stars are assigned to the $P^{(1,2)}$ contribution shown in Fig. 13. The simulation is according to Eq. 30 of text.

**Fig. 15**: Spectral slice from the coarse scan TIFRCARS of Fig. 10b. The simulation is according to Eq. 30.

**Fig. 16**: Two-level logic equivalent to the wiring diagram in Figure 4.

**Fig. 17**: The rotational network controlled with structured pulses for reset (=P), write (=S), process (=P'), and read (G).

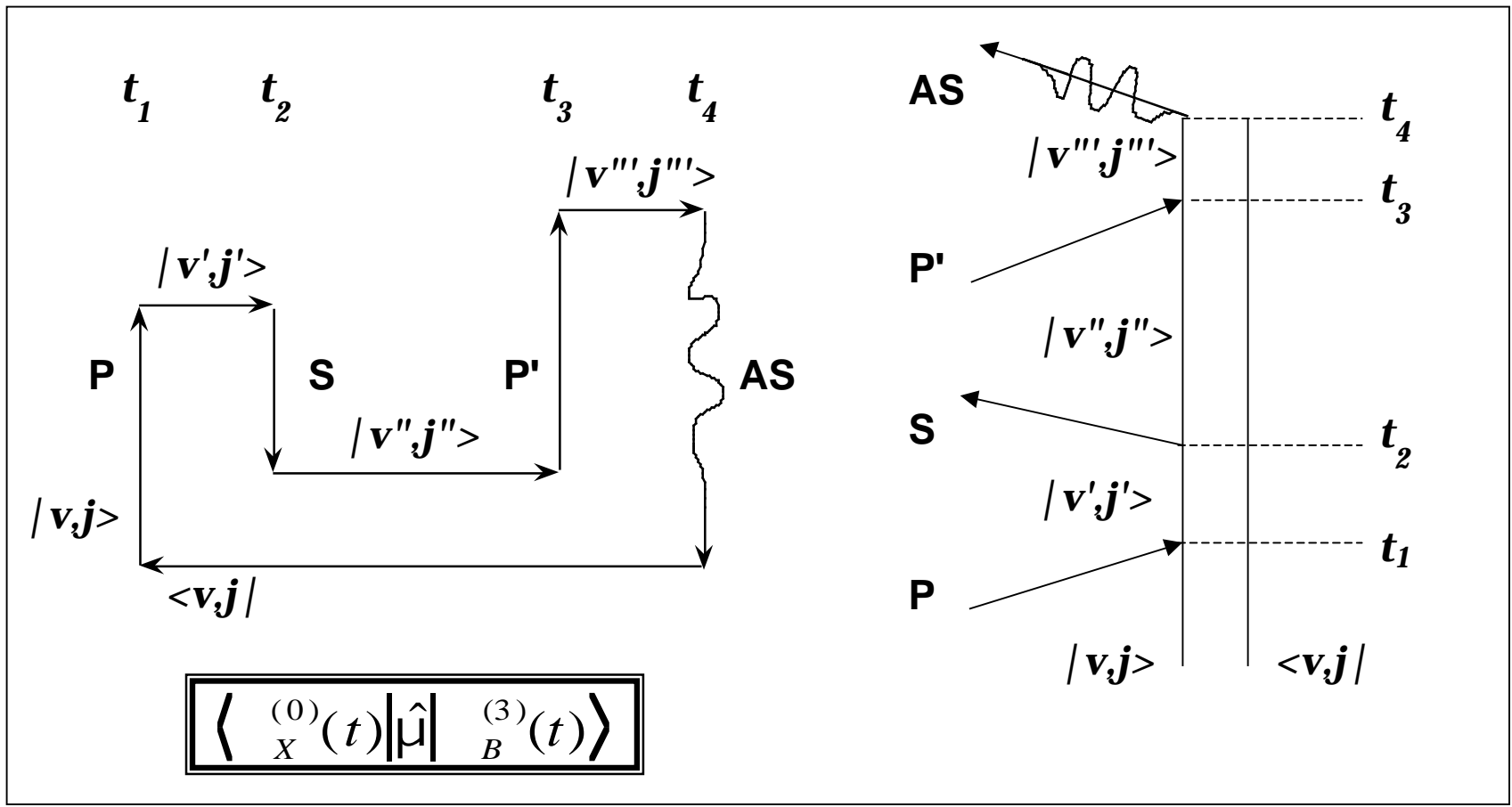

$[t_2 - t_1]$

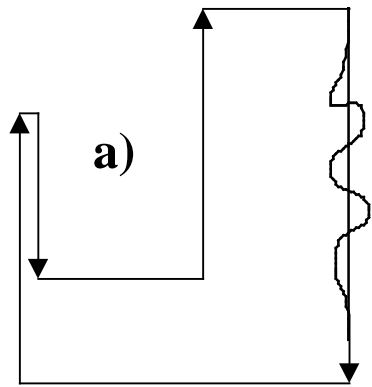 a)  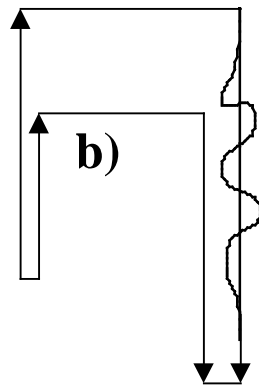 b)  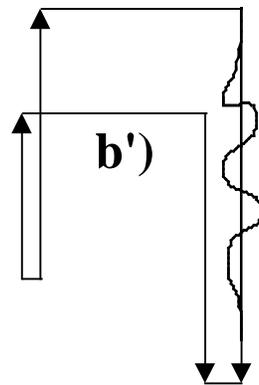 b')  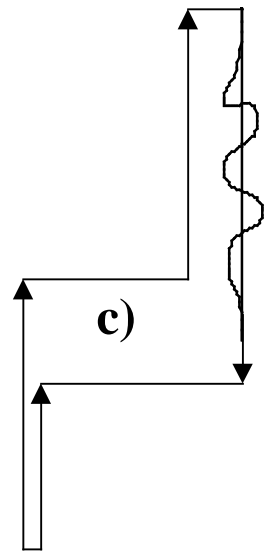 c)

$[t_3 - t_2]$

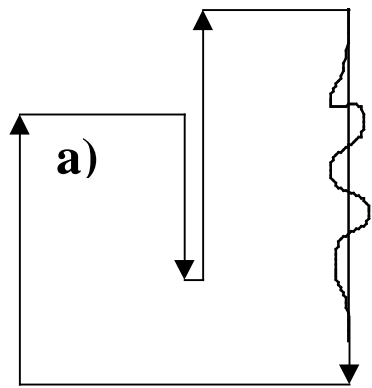 a)  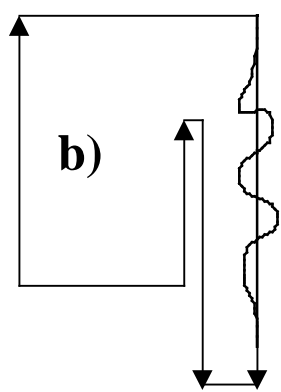 b)  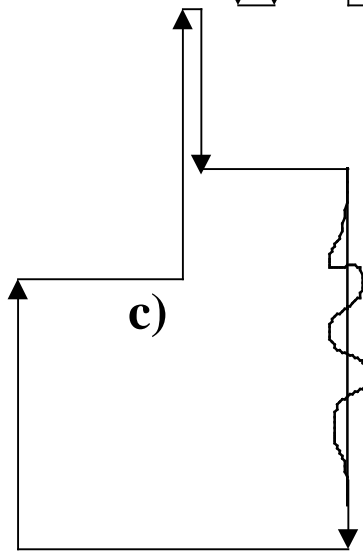 c)

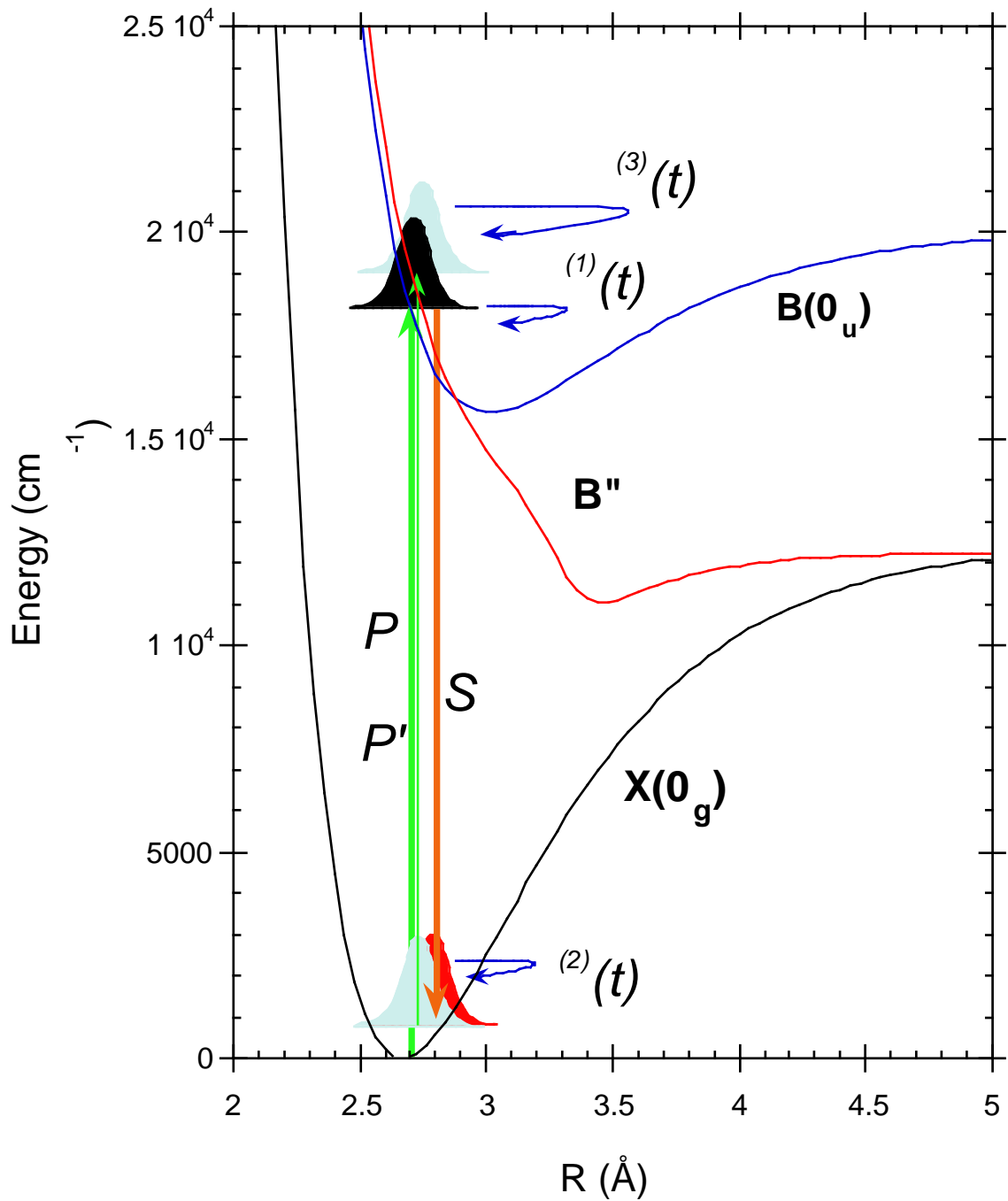

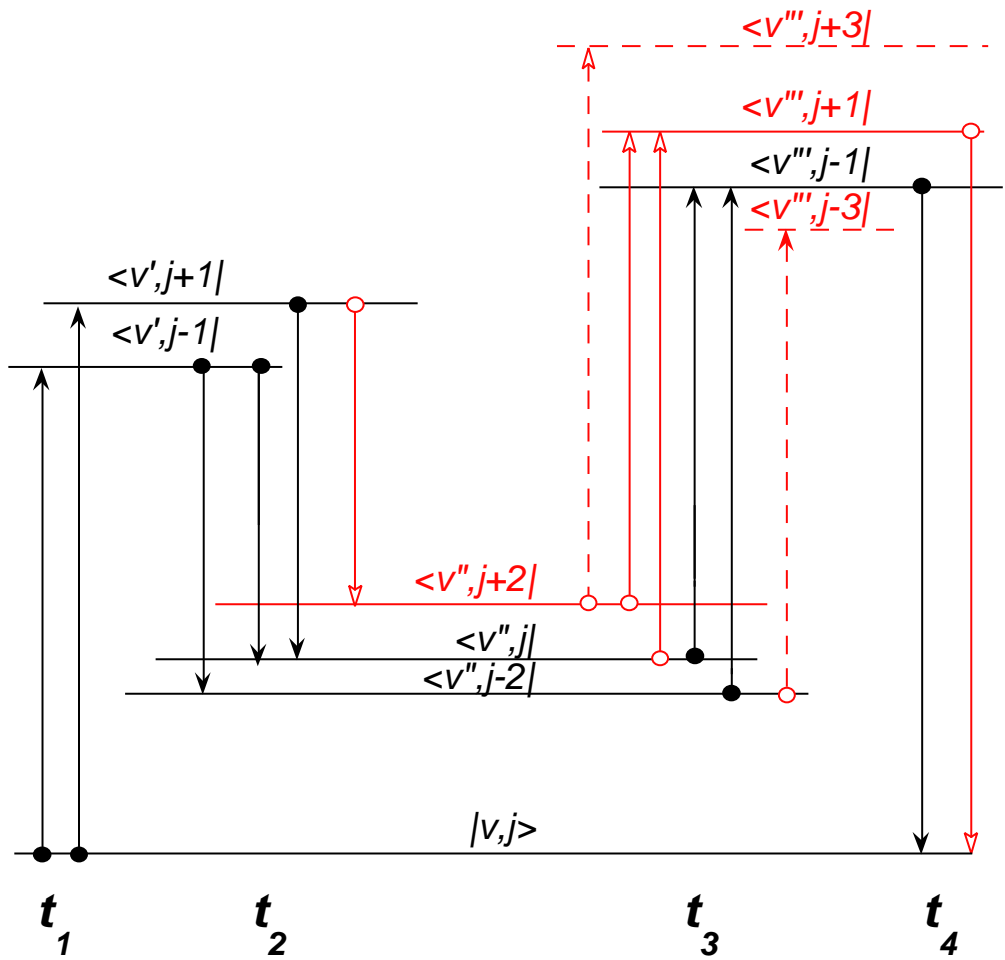

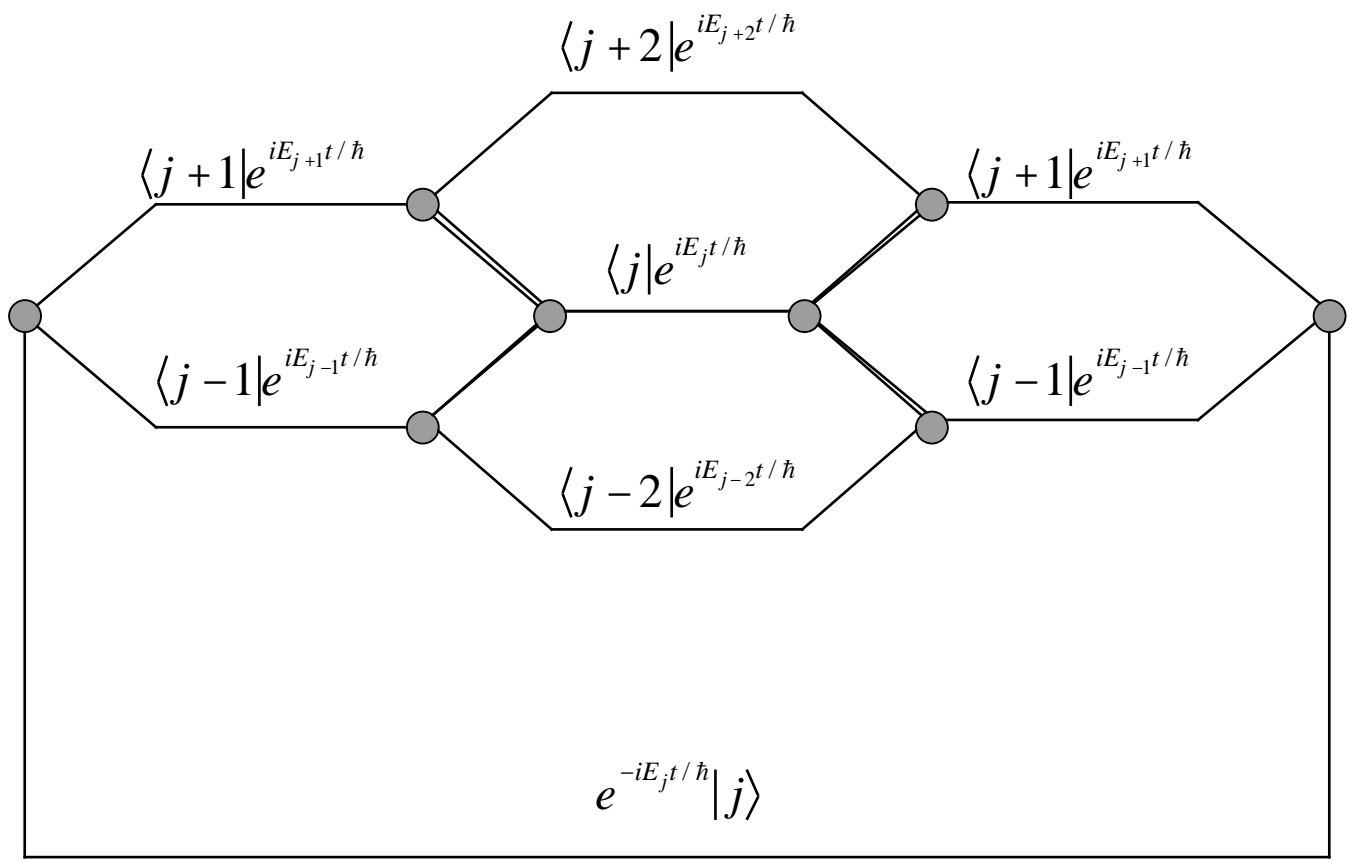

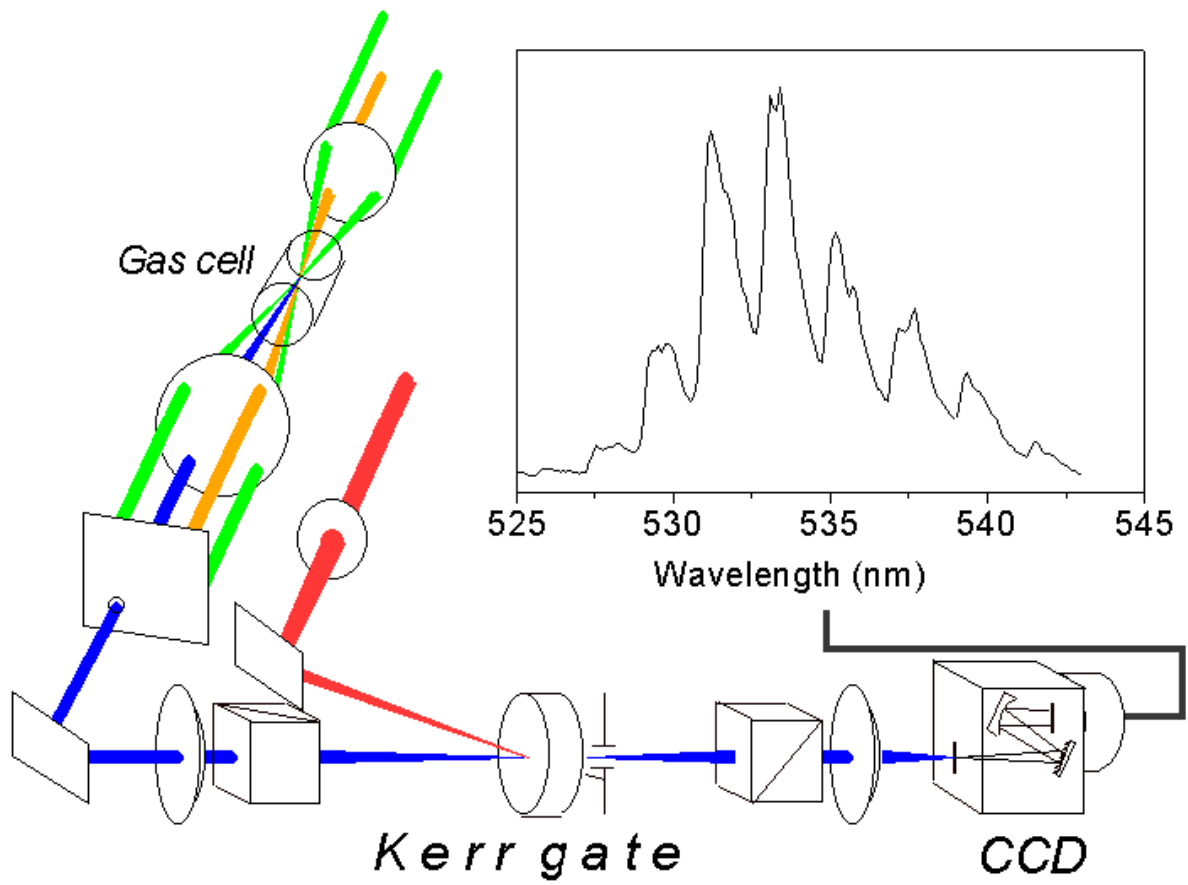

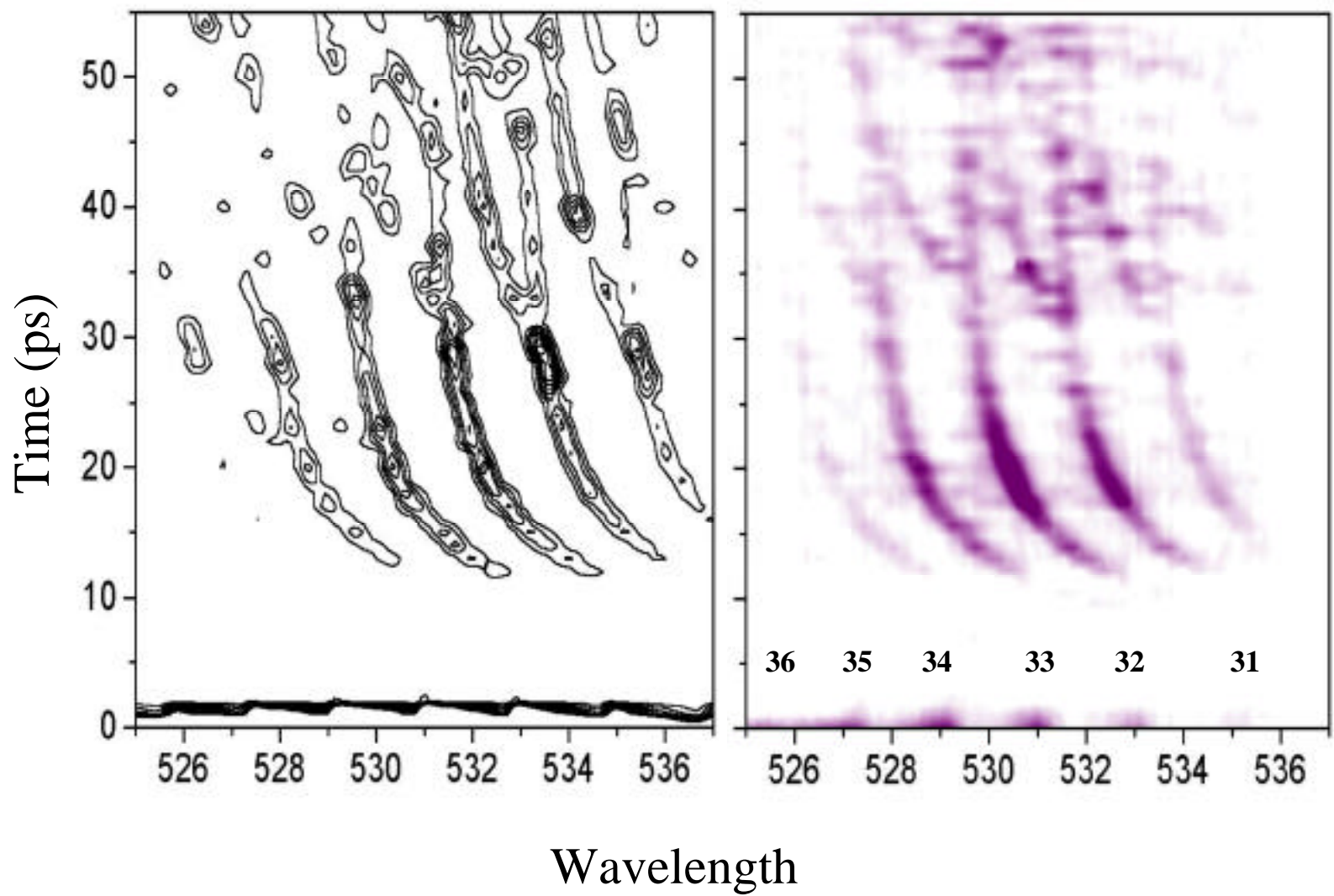

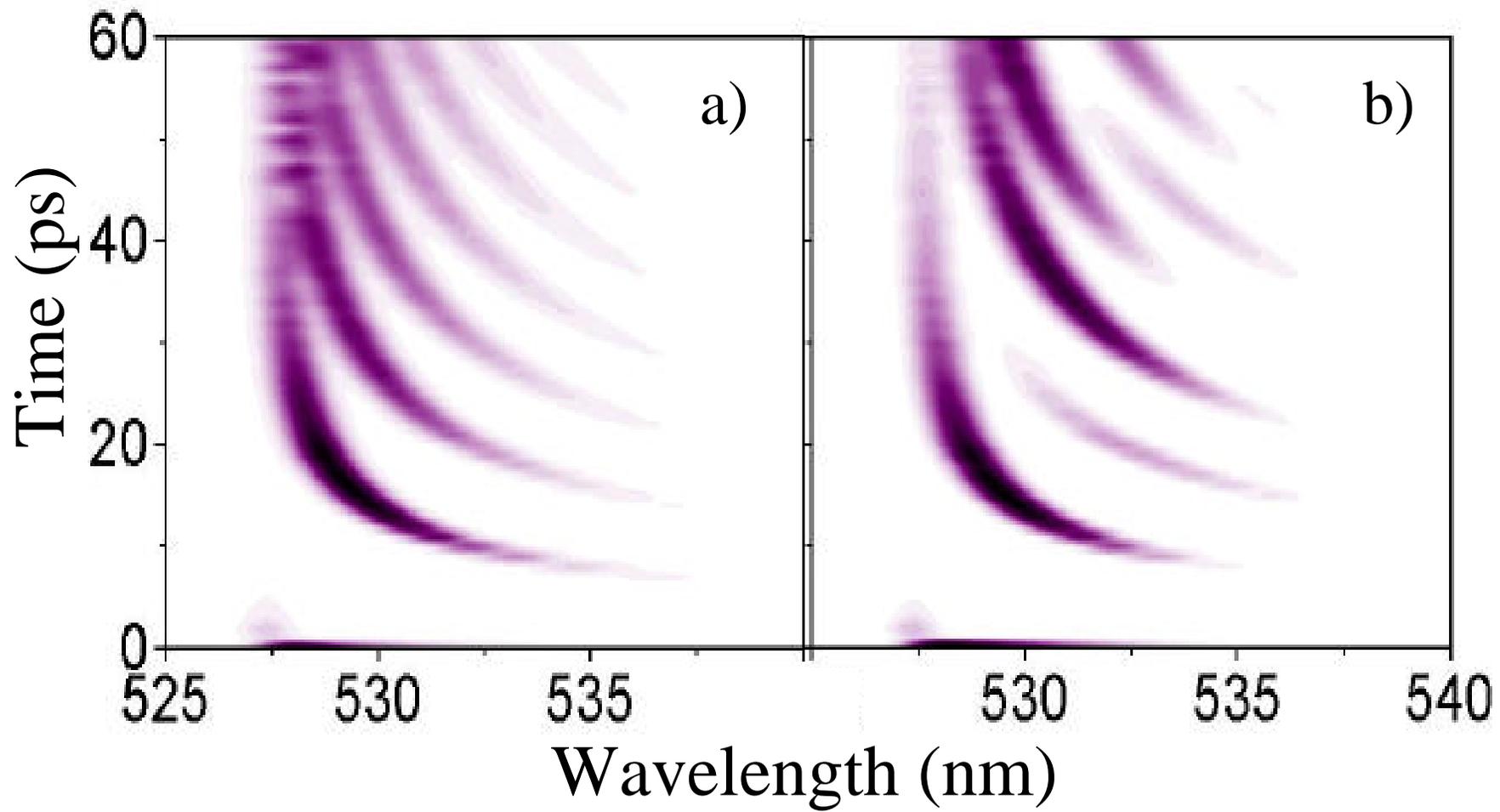

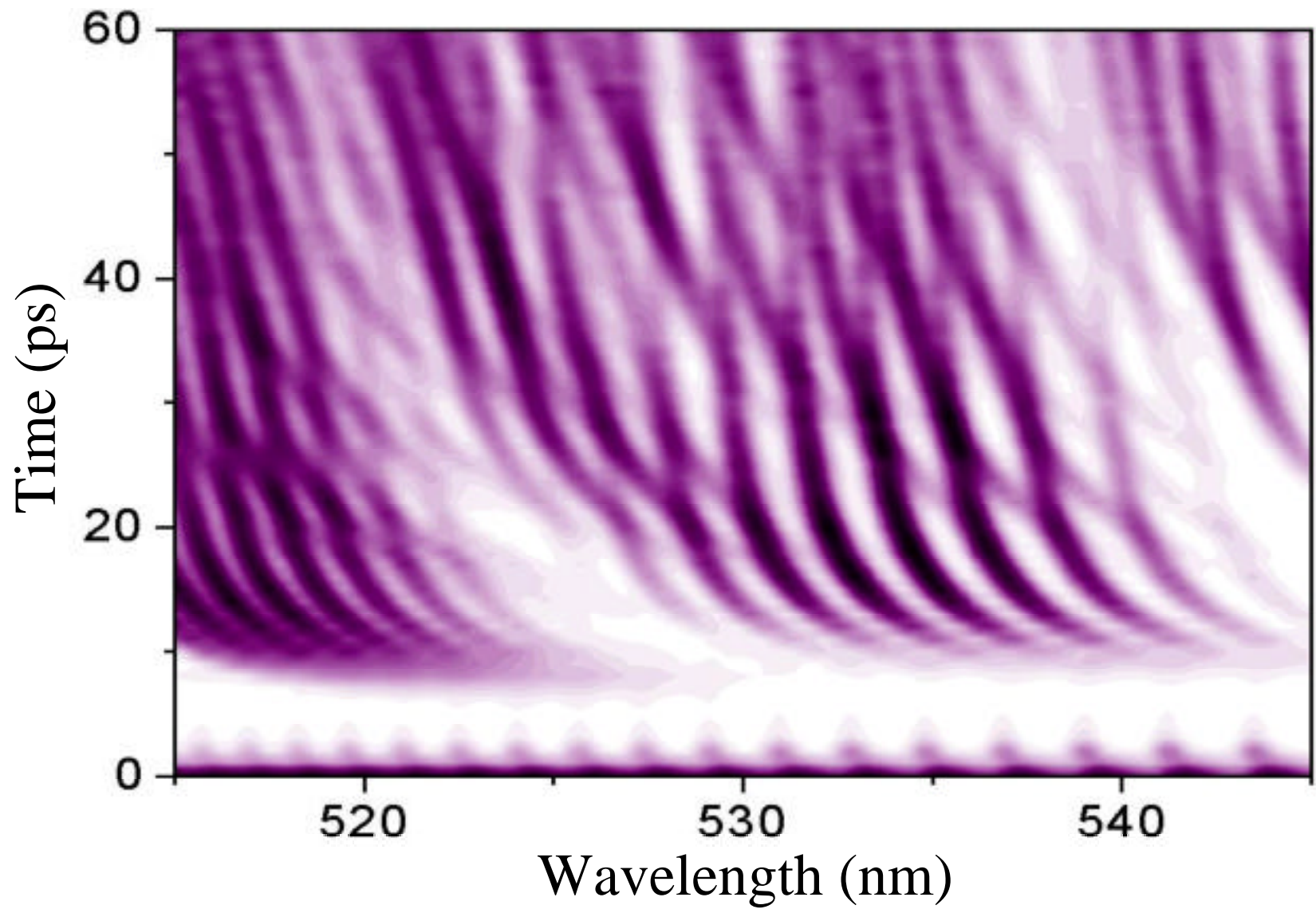

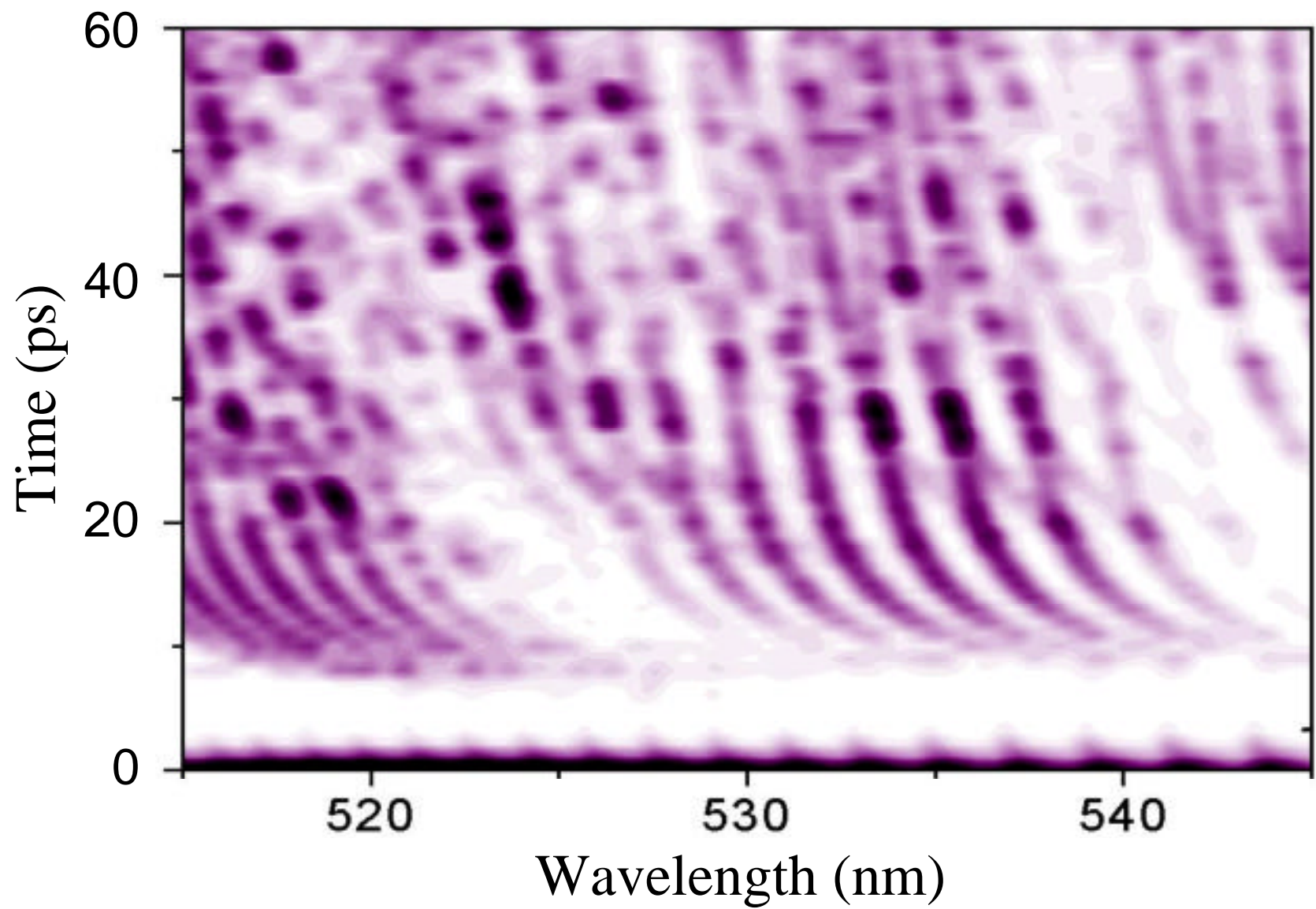

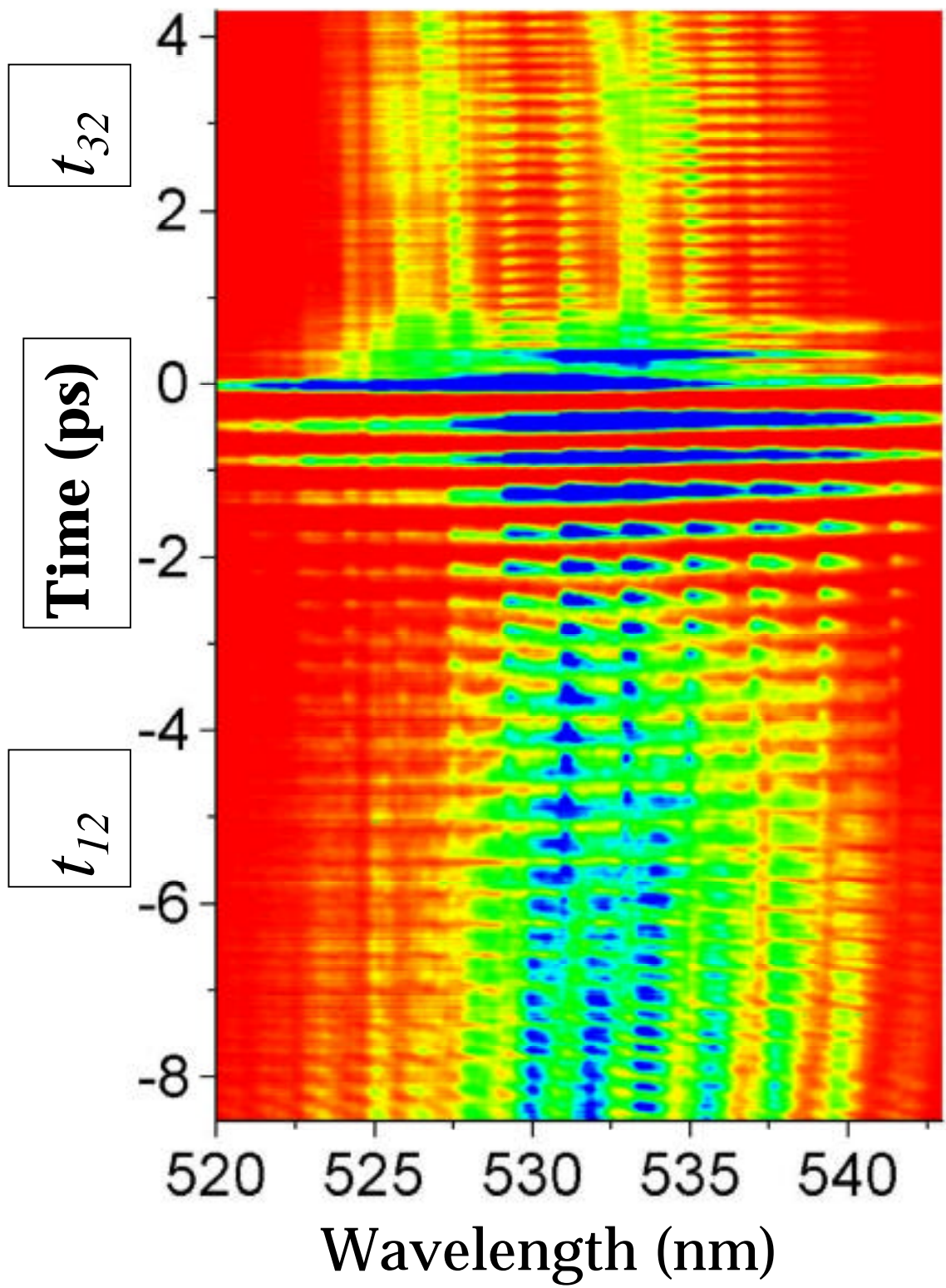

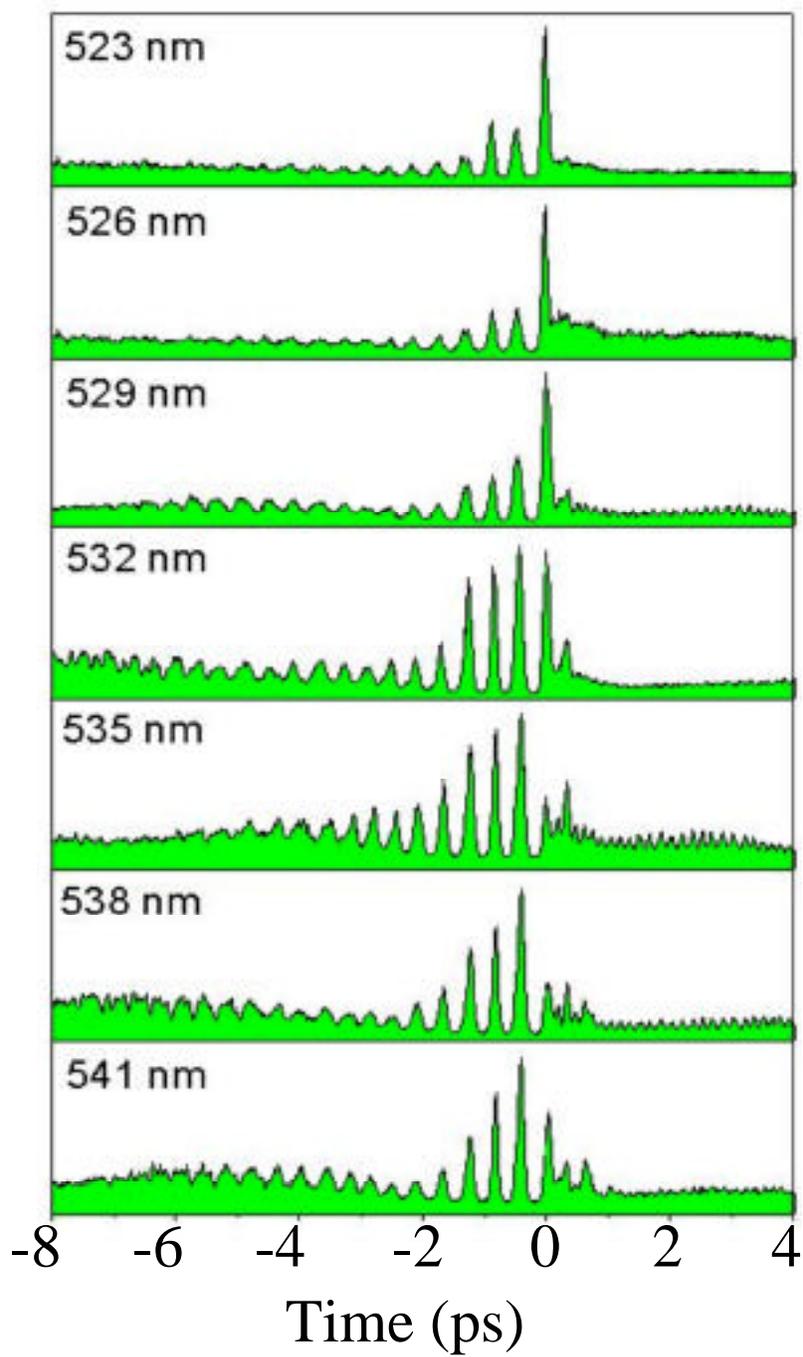

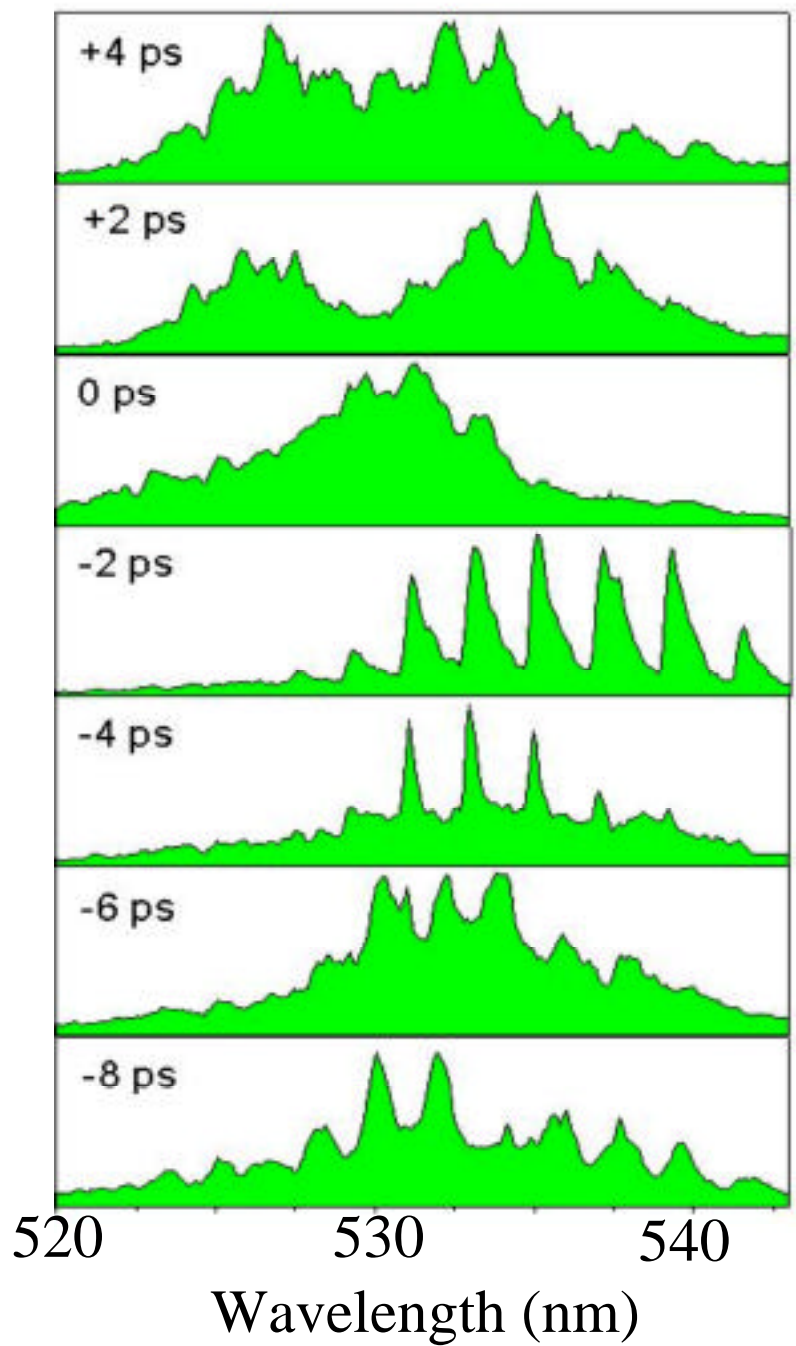

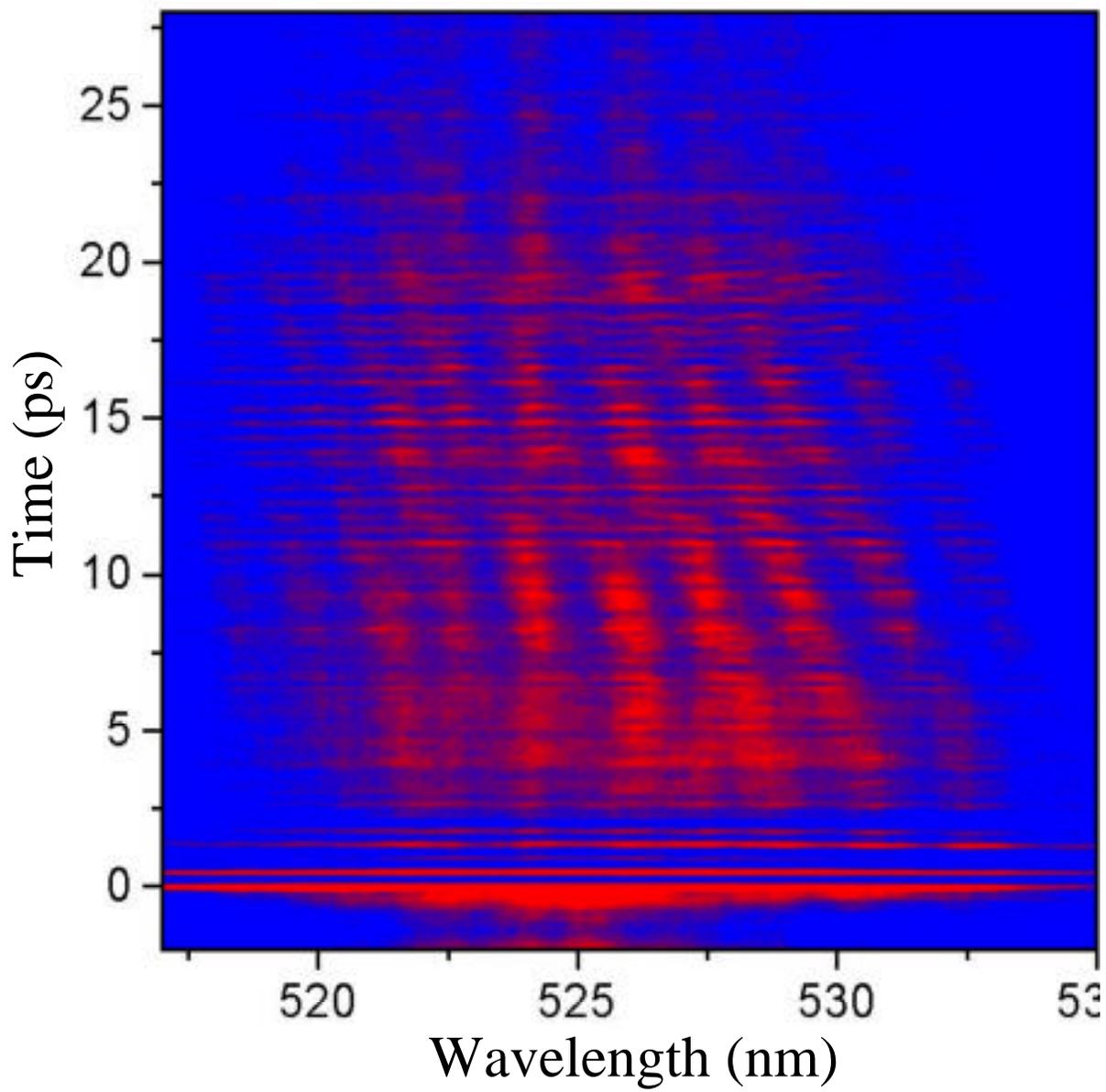

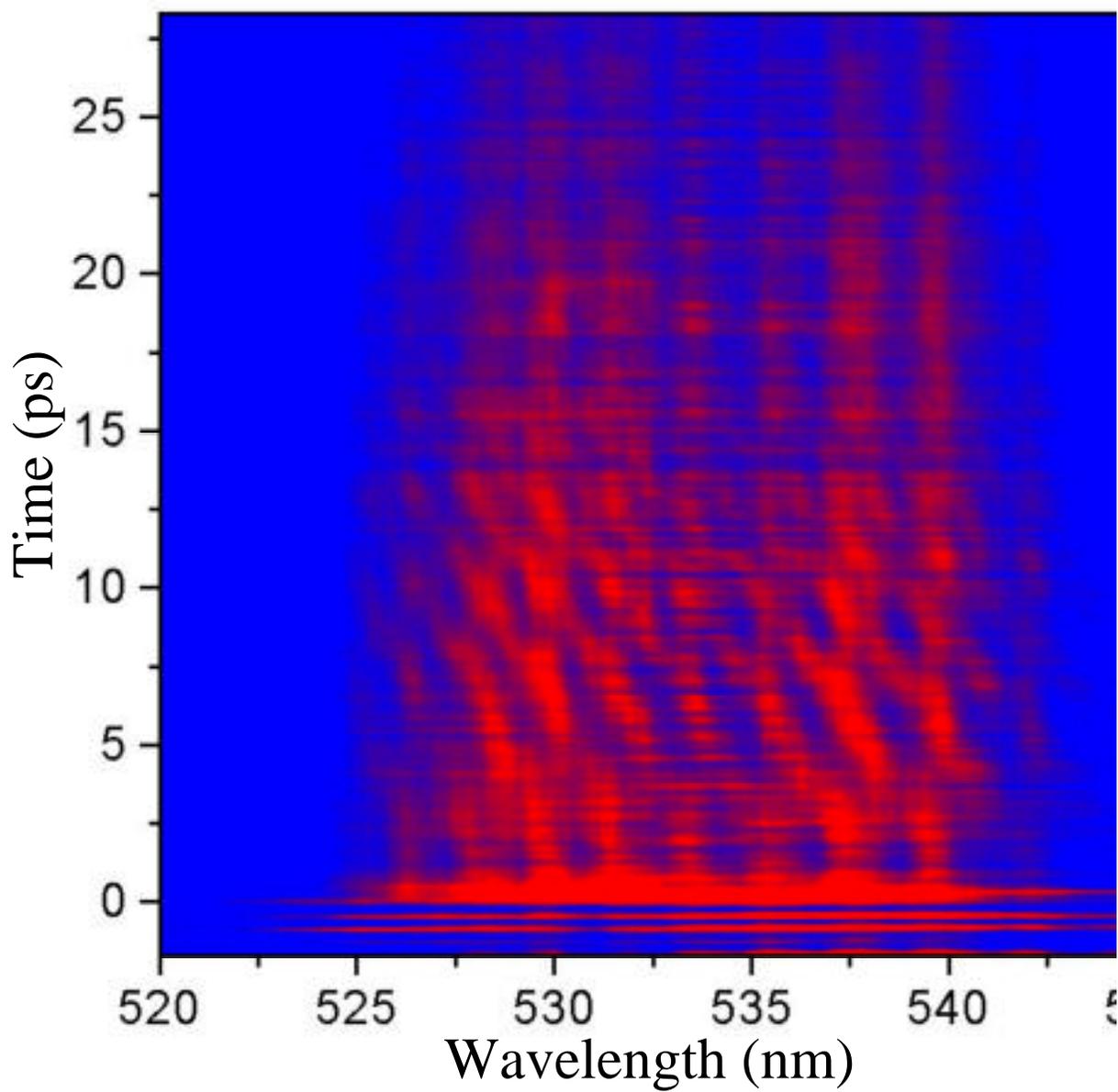

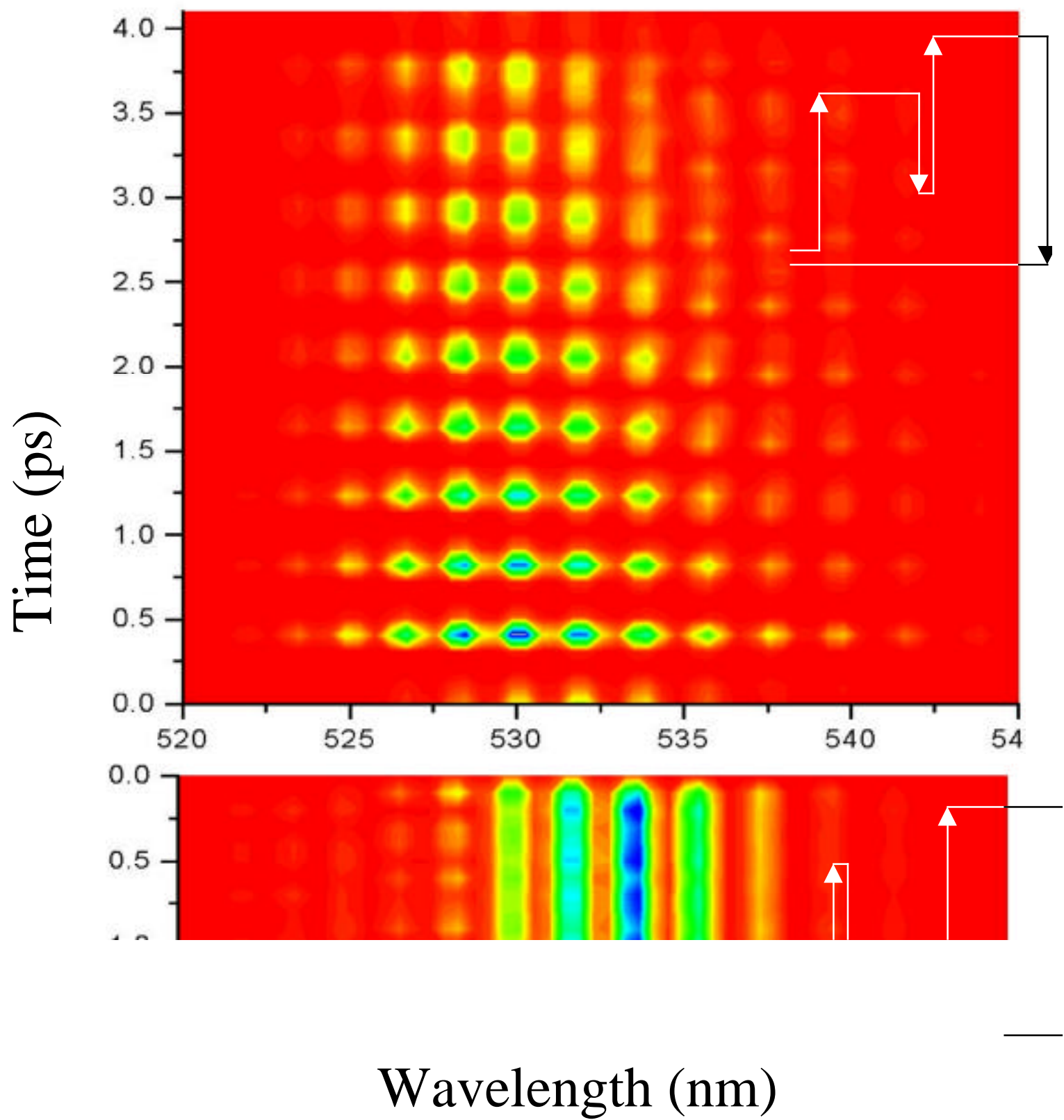

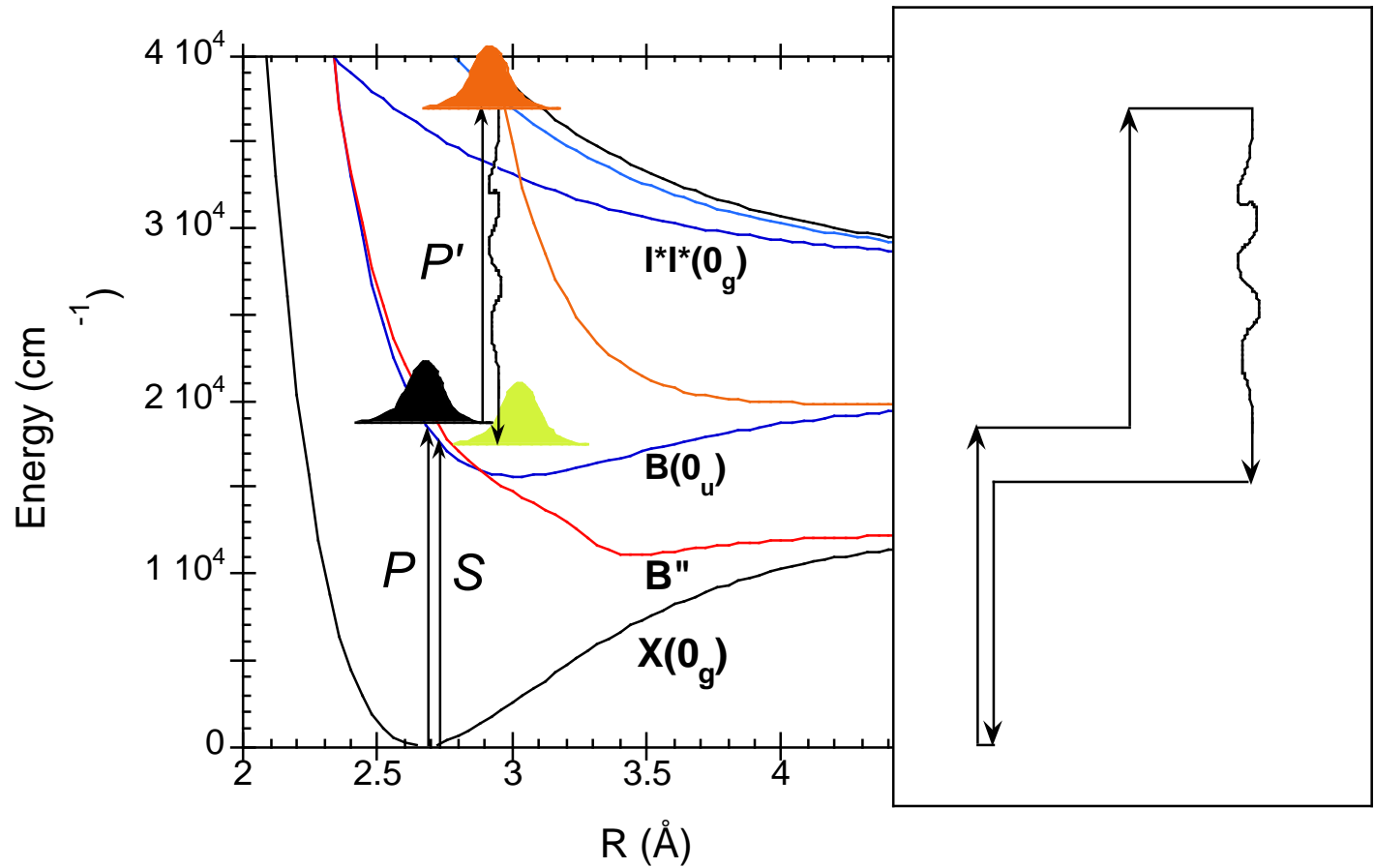

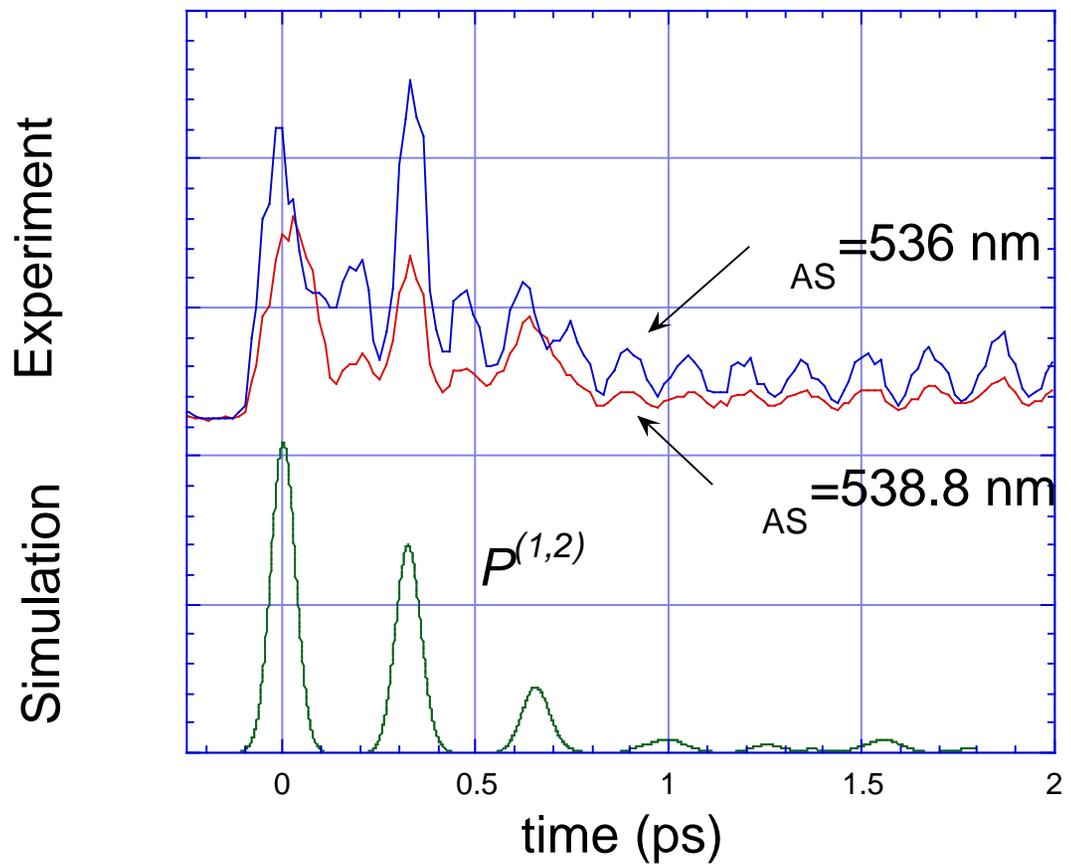

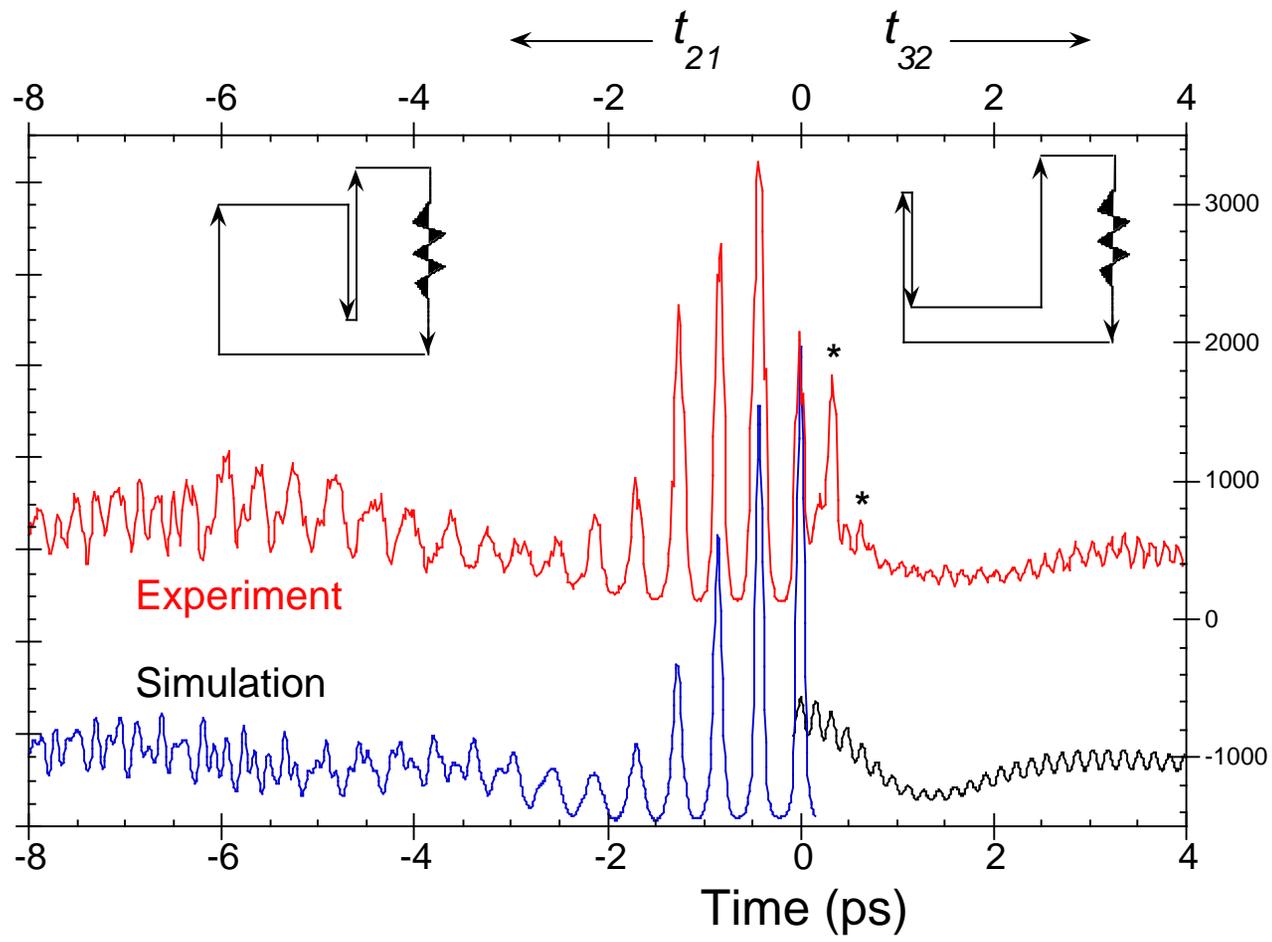

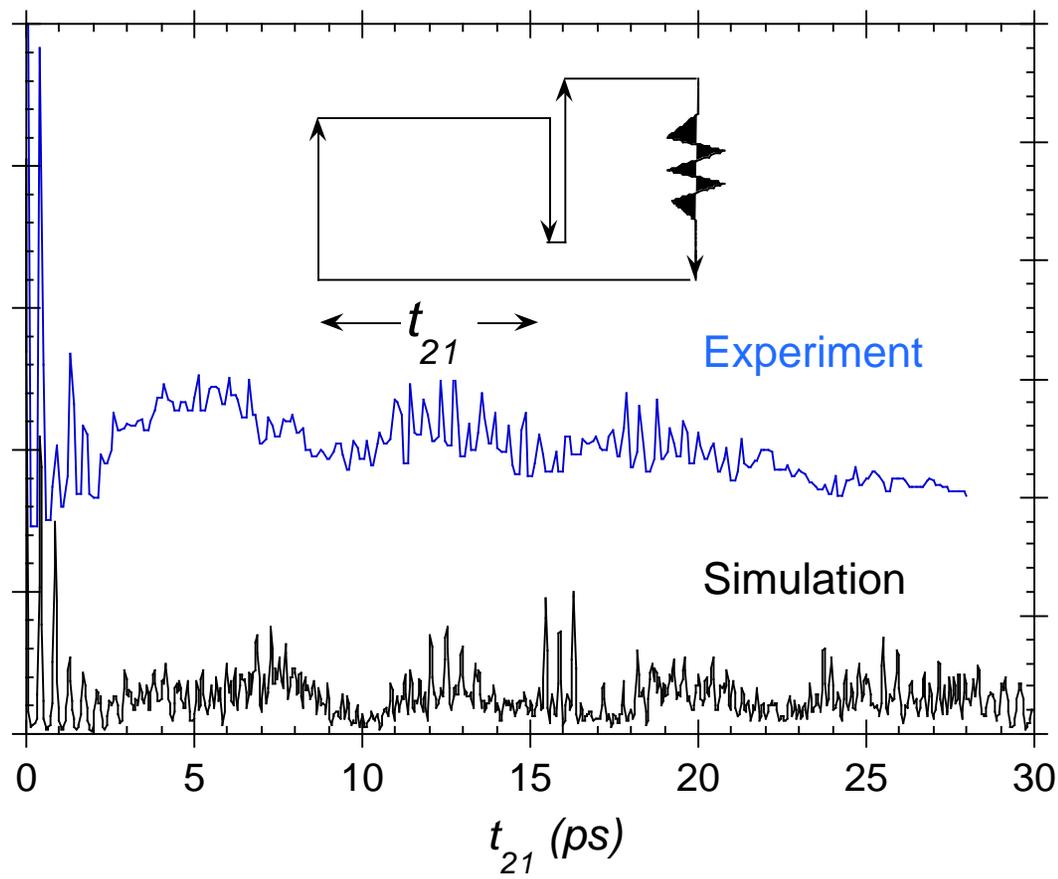

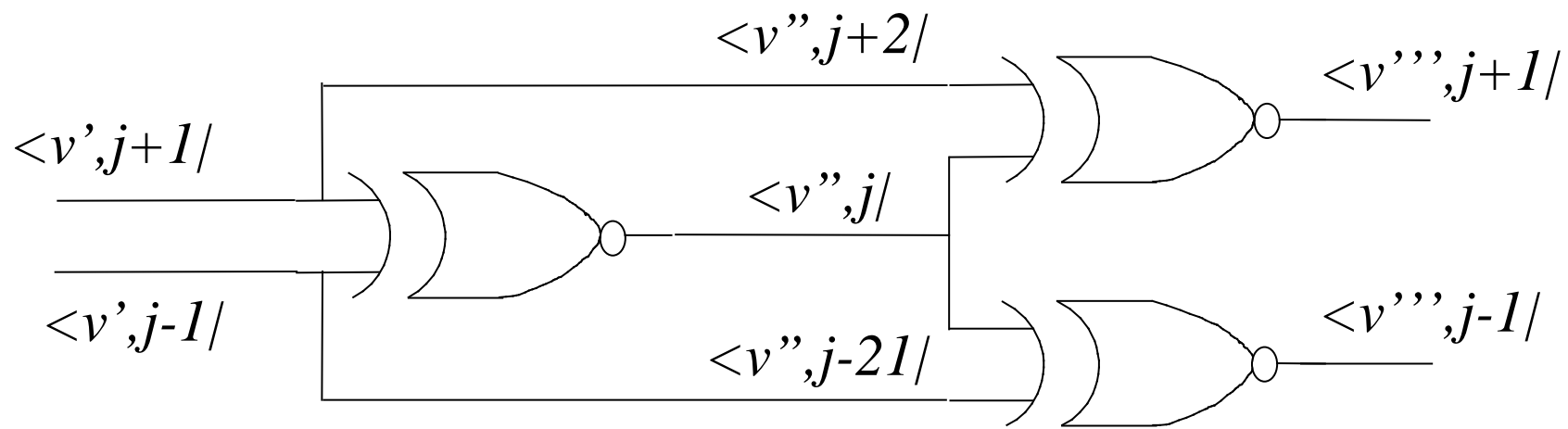

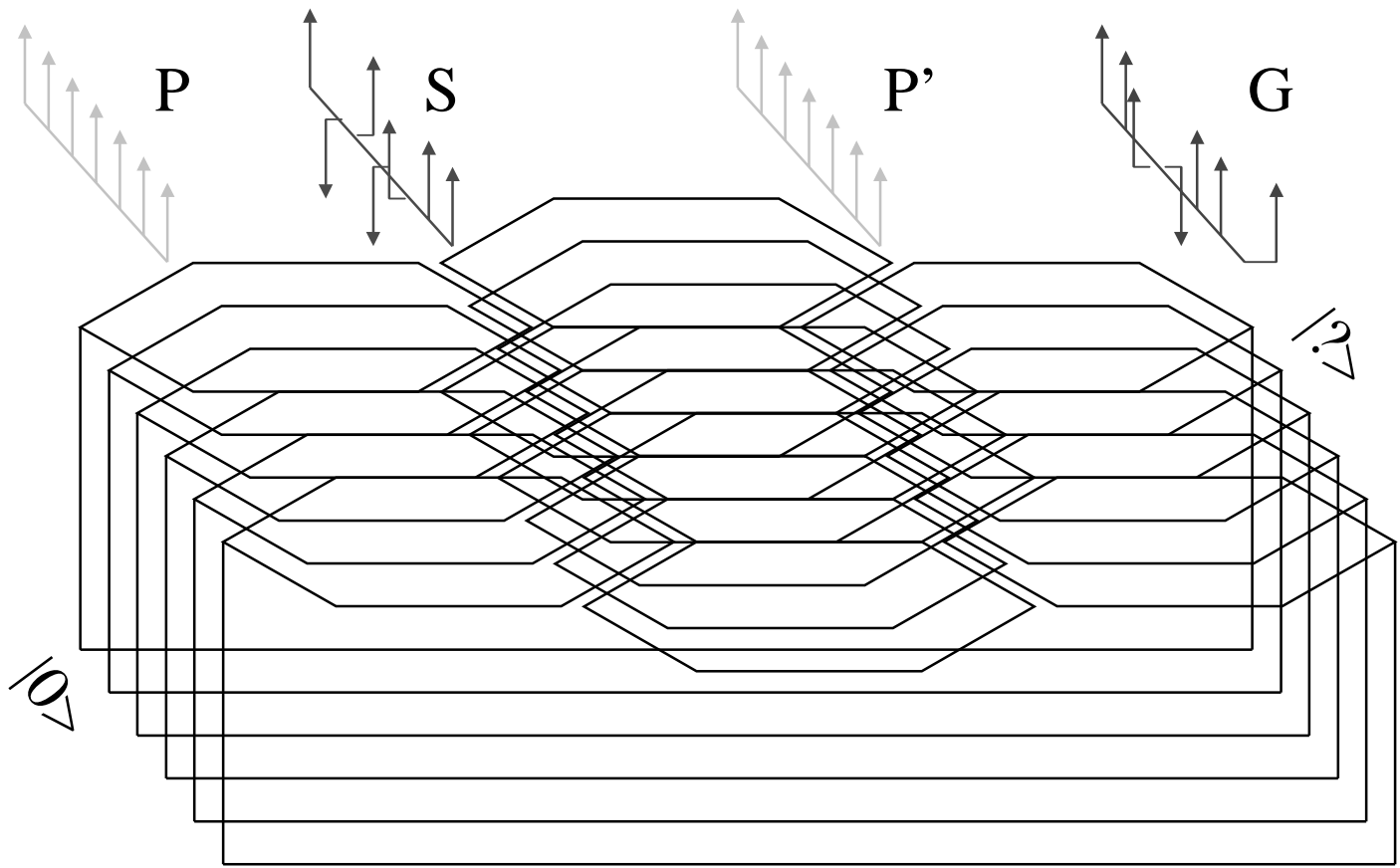